\documentclass[journal,twoside]{asmb}

\begin{document}
\title{Enhancing Physical Layer Security in IoT-Based
RF-FSO Integrated Networks: Multi-RIS Structures
and their Impact on Secure Communication }

\author[1]{Anika Tabassum Biva}
\author[2]{Md. Ibrahim}
\author[1]{A. S. M. Badrudduza}
\author[3]{Imran Shafique Ansari}

\affil[1]{Department of Electronics \& Telecommunication Engineering,RUET}
\affil[2]{Institute of Information and Communication Technology, RUET}
\affil[3]{James Watt School of Engineering, University of Glasgow, Glasgow G12 8QQ, United Kingdom}

\twocolumn[
\begin{@twocolumnfalse}
\maketitle
\begin{abstract}
\section*{Abstract}

Due to their ability to dynamically control the propagation environment, reconfigurable intelligent surfaces (RISs) offer a promising solution to address the challenges of $6$G wireless communication, especially in the context of Internet of Things (IoT) networks. This paper investigates a mixed communication model with multi-RIS-aided radio frequency (RF)-free space optics (FSO) to enhance the performance of IoT applications in complex environments. An eavesdropper is assumed to be present, attempting to intercept confidential information transmitted over the RF link. All RF links are modeled using Rician fading, while the FSO link accounts for Málaga turbulence with pointing errors, capturing real-world propagation conditions. Closed-form analytical expressions are derived for the secrecy outage probability, average secrecy capacity, and effective secrecy throughput in terms of Meijer's G function. To gain further insight, high signal-to-noise approximations of these metrics are also presented. Numerical results highlight the importance of heterodyne detection in mitigating the adverse effects of pointing errors on the FSO link. Moreover, integrating a multi-RIS structure into the proposed model significantly increases secrecy performance, achieving up to a $47.67\%$ improvement in SOP compared to conventional methods. Finally, the derived analytical results are validated through Monte Carlo simulations.
\end{abstract}
\begin{IEEEkeywords}
\section*{Keywords} 
RIS, dual-hop network, FSO communication, pointing error, Málaga turbulence, Rician fading.
\end{IEEEkeywords}
\end{@twocolumnfalse}
]

\section{Introduction}

\subsection{Background}

Reconfigurable intelligent surfaces (RISs) have emerged as a pivotal player in the quest for secure Internet of Things (IoT) connectivity and enhanced communication standards \cite{shi2023outage,basharat2022reconfigurable,RAKIB2024}. Comprising cost-effective and energy-efficient modules, RISs serve as a beacon of innovation by intelligently manipulating incoming signals \cite{han2020cooperative,tasci2022new}. By manipulating these shifts, RISs can precisely control the direction of transmitted beams, ensuring signals reach their destination flawlessly \cite{peng2022performance,rakib2024ris}.
Moreover, through strategic relay deployment, dual-hop radio frequency (RF) - free space optics (FSO) mixed models seamlessly integrate the strengths of both RF and FSO communication technologies. While RF stands as a robust and versatile option for next-generation wireless communications, FSO represents a higher-licensed optical spectrum line-of-sight (LOS) technology. However, RISs act as a backbone in this IoT network, supporting both RF and FSO transmissions to mitigate signal blockage issues and enhance overall functionality.

\subsection{Literature Review}
Recent advancements in mixed communication systems, particularly those integrating FSO and RF technologies, have significantly contributed to enhancing wireless communication capabilities. For instance, the performance of dual-hop RF-FSO systems was extensively analyzed, revealing critical insights into outage probability (OP) and bit error rates (BER) under various fading conditions and detection techniques \cite{sun2021performance}. In the context of space-air-ground integrated networks, unmanned aerial vehicle (UAV)-assisted systems were shown to improve reliability and capacity, with detailed analyses providing closed-form expressions for OP and average BER \cite{qu2022uav}. Further studies explored fixed-gain relaying in FSO-RF systems, highlighting the impact of Fisher–Snedecor ($\mathcal{F}$) and $\kappa-\mu$ shadowed fading on system performance \cite{ding2022performance}. Additionally, the influence of co-channel interference on mixed RF-FSO systems was evaluated, providing new closed-form expressions for performance metrics in both fixed-gain and variable-gain relaying scenarios \cite{ding2023joint}. Theoretical frameworks for dual-hop mixed FSO-RF systems were developed, addressing various performance aspects, including OP and ergodic capacity (EC) \cite{ashrafzadeh2019framework}. Two-way decode-and-forward (DF) relaying systems with co-channel interference were also investigated, with findings demonstrating the effects of Nakagami-$m$ fading and the double generalized Gamma scintillation model on performance \cite{tonk2020mixed}. Research on higher-order quadrature amplitude modulation schemes revealed the impact of outdated channel state information and pointing errors, contributing to a deeper understanding of system performance under various conditions \cite{singya2020performance}. High-throughput satellite systems utilizing mixed FSO-RF transmission were studied, where the authors derived the analytical expression of EC, incorporating techniques such as beamforming to mitigate atmospheric turbulence and maximize capacity \cite{kong2021ergodic}. The performance of satellite-terrestrial systems was analyzed, providing precise closed-form expressions for various performance metrics \cite{sun2024performance}. Lastly, hybrid FSO/RF-THz relay systems were proposed to overcome the limitations of traditional RF systems, demonstrating improved performance through adaptive combining schemes \cite{liang2024performance}.

Due to its ability to enhance connectivity, energy efficiency, spectrum utilization, and security, RISs technology offers promising opportunities for improving IoT networks, making it a compelling research area for the IoT community \cite{khalid2023reconfigurable,bhowal2022ris,zhang2024performance}. Recently, RISs have emerged as a promising technology for enhancing single-hop communication systems. For instance, in \cite{atapattu2020reconfigurable}, the authors investigated RIS-assisted two-way communications and derived exact closed-form expressions for OP and spectral efficiency. This research demonstrated that RISs with multiple elements could substantially enhance system performance in Rayleigh fading environments. Building upon this, the authors of \cite{guo2020outage} examined IRS-assisted systems and obtained analytical expressions for OP and optimal phase shifts. Furthermore, \cite{selimis2021performance} analyzed the performance of RISs in Nakagami-$m$ fading channels, providing expressions for OP, BER, and EC. In another related study, the authors of \cite{basu2024performance} investigated RIS-assisted index modulation and assessed its performance using space-shift keying (SSK) and spatial modulation in Rician fading channels. The analysis demonstrated the positive impact of RIS, emphasizing the performance gains achieved with RIS-assisted systems. Recently, the authors of \cite{zhu2023ris} examined full-duplex SSK systems, presenting a detailed analysis of ABER under intelligent and blind phase-shift schemes. The study concluded that RIS-FD-SSK outperforms conventional half-duplex systems.
In the context of dual-hop RIS-assisted works, several studies have thoroughly examined the impact of RIS on enhancing communication systems, particularly in mixed RF-FSO networks. A notable study in \cite{9057633} investigated a dual-hop RIS-assisted communication system, where RIS plays a crucial role in improving coverage and system performance by addressing atmospheric turbulence and pointing errors. In a similar vein, research on multi-hop RIS-assisted UAV communications in \cite{abualhayja2023exploiting} demonstrated that strategically deploying RIS in UAV communication systems significantly enhances signal-to-noise ratio (SNR) and minimizes OP. This is particularly important when ideal LOS conditions are inconsistent due to the dynamic mobility of the UAV. Moreover, the work in \cite{9424709} on mixed RF-FSO relay networks compares two configurations: an RIS-equipped RF source and an RIS-aided RF source. This study revealed that at high SNR, system performance was dominated by the worst communication hop and that RIS-equipped sources outperform RIS-aided sources in terms of both diversity and coding gains. Lastly, another work in \cite{aldababsa2023multiple} focused on multiple RISs-aided networks with opportunistic RIS scheduling in a dual-hop scenario where the study emphasized the advantage of deploying multiple RISs to achieve a higher diversity order and improved system performance.

In the era of $6$G, the increasing vulnerability of wireless networks to eavesdropping and security threats has made secure communication a crucial priority \cite{10697101}. Wireless communication faces challenges in protecting information privacy due to its inherent vulnerability to security threats \cite{badrudduza2021security}. Within this field, physical layer security (PLS) has emerged as a promising alternative to conventional encryption approaches \cite{shakir2021physical, mitev2023physical, 10460296}. Recent studies have extensively investigated the secrecy performance of combined RF-FSO systems \cite{saber2024security, 10058969, zhang2024joint}. Furthermore, RISs also underscore their transformative potential in single-hop networks, particularly in enhancing security performance. For instance, one study in \cite{yang2020secrecy} presented analytical expressions for secrecy outage probability (SOP) and validated the effectiveness of RIS in improving secrecy performance. Additionally, the performance of RIS with spatially random eavesdroppers over Rician channels was analyzed in \cite{shi2024secrecy}. In the context of smart grid communications, RIS was proposed as a means to enhance PLS \cite{kaveh2023secrecy}. Another study examined the impact of RIS on STAR-RIS non-orthogonal multiple access (NOMA) networks, focusing on secrecy performance in the presence of residual hardware impairments \cite{li2022enhancing}. Recently, the authors examined the performance of RIS-assisted index modulation systems over Rician fading channels, focusing on full-duplex systems with various modulation schemes in \cite{zhang2022secrecy}.
Furthermore, the performance limits of multi-hop RIS-assisted UAV communications were analyzed in \cite{yadav2023secrecy}, demonstrating the strategic RIS placement and element numbers can enhance wireless communication. The authors of \cite{hoang2023secrecy} investigated the RIS-aided security performance where they proposed an alternating optimization for beam-forming and RIS-reflecting vectors, showing that the double-RIS scheme significantly outperforms the single-RIS approach in security. Lastly, the impact of co-channel interference on RIS-assisted networks was investigated, focusing on multiple eavesdropping attempts and their effects \cite{ruku2024effects}. Although many recent works have investigated the secrecy performance of RIS-aided systems in single-hop networks, very few studies have explored secrecy performance in the RF-FSO mixed networks. In \cite{wang2023uplink}, the authors analyzed the secrecy performance of RIS-based heterogeneous networks, deriving closed-form expressions for SOP in multi-user scenarios. They highlighted the significant impact of RIS elements, atmospheric turbulence, and pointing errors in secrecy. Similarly, \cite{10313311} investigated RIS-aided mixed networks, exploring various eavesdropping scenarios and evaluating metrics like average secrecy capacity (ASC) and effective secrecy throughput (EST) where the authors emphasized the role of fading, turbulence, and detection techniques in enhancing secrecy. Furthermore, \cite{ahmed2023enhancing} focused on improving PLS performance in RIS-aided RF-FSO systems, providing analytical expressions for various secrecy metrics and analyzing the impact of simultaneous eavesdropping on both RF and FSO links. Finally, \cite{zhuang2022secrecy} addressed the challenge of imperfect channel state information in a NOMA network, deriving SOP expressions under Gamma-Gamma (GG) distributions for FSO links and Rayleigh fading for RF links.

\subsection{Motivation and Contributions}
Due to the combined strengths of RF and FSO technologies, mixed networks provide a strategic solution for enhancing security performance in future wireless networks. Moreover, RISs are particularly beneficial in RF-FSO mixed networks as they enhance signal strength, coverage, and resilience, even in adverse conditions. Furthermore, RIS selection in RF networks significantly influences overall system performance, as strategically placed RISs can improve signal strength, providing improved security and reliability through enhanced diversity. Although considerable research has been conducted on RIS-assisted single-hop networks to investigate secrecy performance, the focus on mixed networks remains limited. Dual-hop systems, particularly those involving multiple RISs, offer an opportunity to enhance security further by introducing diversity, which is important for improving secrecy performance in complex wireless environments. Despite the benefits of such configurations, existing literature has yet to fully explore the potential of multiple RISs for secure communications in dual-hop RF-FSO networks, leaving a significant gap that our study seeks to address. In this paper, a multi-RIS-aided RF-FSO mixed network is analyzed where a malicious eavesdropper attempts to intercept sensitive information transmitted over the RF link. To bridge the considerable distance between the source and destination, a strategically located relay is employed to facilitate uninterrupted communication. Additionally, the presence of obstacles prevents a direct connection between the source and relay, necessitating the deployment of an RIS to assist in signal reflection and amplification. The RF link is modeled as a Rician fading channel, accurately capturing scenarios with a dominant LOS component and scattered multi-path signals \cite{10313311}, while the FSO link is modeled using the Málaga turbulence model due to its accuracy and flexibility in representing the impact of atmospheric turbulence on optical signal propagation \cite{10058969}. However, the proposed model provides valuable insights into the potential of employing multiple RIS to increase the secrecy performance of RF-FSO networks. However, the major contributions of this research paper are mentioned as follows:
\begin{itemize}
    \item We derive closed-form expressions for the cumulative distribution function (CDF) of the multi-RIS-assisted RF-FSO network, utilizing the CDFs and PDFs of the individual links. While previous studies have examined the secrecy performance of RIS-aided networks, to the best of our knowledge, this is the first work to consider a multi-RIS structure instead of a single RIS unit, thereby analyzing a more realistic system model.

    \item Utilizing the derived CDF of our proposed model, we analytically derive closed-form expressions for the lower bounds of SOP, ASC, and EST. These expressions are then evaluated numerically for specific system configurations. To validate the accuracy of our analytical results, we conduct extensive Monte Carlo simulations. The strong agreement between the analytical and simulation results confirms the reliability of our analysis.

    \item To enhance the applicability of our analysis, we provide valuable insights into the design of secure multi-RIS-assisted RF-FSO mixed networks. Our analysis considers the key impairments and characteristics of both RF and FSO links, ensuring a more realistic representation. We investigate the effects of fading parameters, RIS elements, and the number of RISs for RF links, as well as the impacts of atmospheric turbulence, detection techniques, and pointing error conditions for FSO links.

    \item A comparison with existing methods demonstrates that the proposed RIS selection strategy achieves a 47.67\% improvement in SOP under strong turbulence conditions at an average SNR of 10 dB.

\end{itemize}

\subsection{Organization}
The remainder of the paper is organized as follows: Section \ref{sys} presents the proposed system model and the statistical analysis results for the PDF and CDF of each link. Section \ref{pm} derives the performance metrics of SOP, ASC, and EST. Section \ref{nr} presents the numerical results and Monte Carlo simulations. Finally, Section \ref{con} concludes the paper with a summary.

\section{System Model and Problem Formulation} \label{sys}
\begin{figure}[!h]
    \centering
   \includegraphics[width=0.5\textwidth]{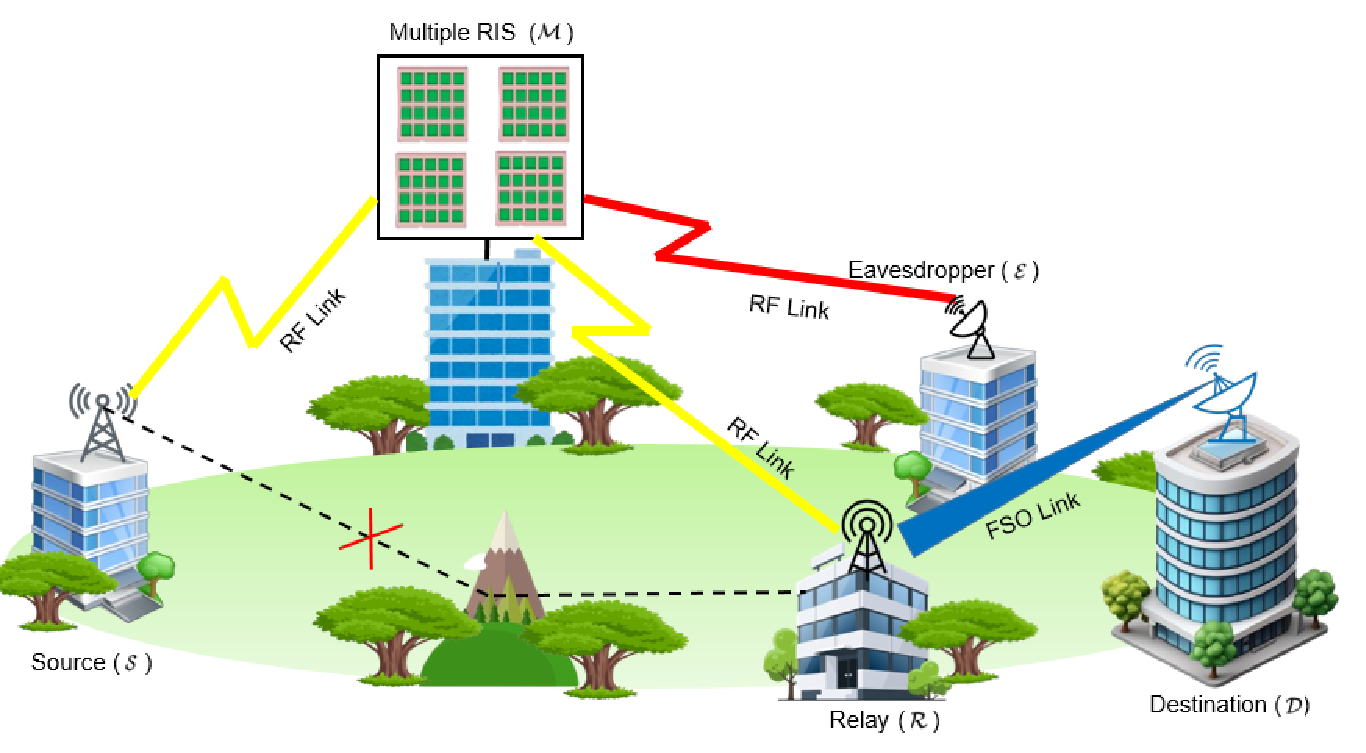}
    \caption{System model of a multi RIS- assisted over mixed RF-FSO communication}
    \label{Fig:system}
\end{figure}
A multi-RIS assisted  RF-FSO mixed communication system consists of a stable source ($\mathcal{S}$), a relay ($\mathcal{R}$), multiple RISs ($\mathcal{M}$) and a destination ($\mathcal{D}$), is shown in Fig. \ref{Fig:system}. In the proposed model, information is transmitted over two hops. In the first hop, we assume a scenario in which a base station in an urban area needs to transmit data to a relay station, located on top of a building. Despite some environmental obstacles, $\mathcal{S}$ and $\mathcal{R}$ are not directly connected. As a result, the incoming signal from $\mathcal{S}$ is first transmitted to $\mathcal{M}$, then transformed into $\mathcal{R}$. In a dense urban environment, the base station communicates with $\mathcal{R}$ through RISs mounted on buildings to overcome obstacles such as tall structures or complex terrain, which would otherwise block the RF signal.

\noindent Assume that all $\mathcal{S}-\mathcal{M}$ links are subjected to the Rician distribution. Each RIS, $\{RIS_{m}\}^{M}_{m=1}$, is equipped with $N$ reflecting elements where the channel vector is expressed by $\boldsymbol{h}_{m}$, where $\boldsymbol{h}_{m}$=$\left [ h^{(1)}_{m},\cdots, h^{(i)}_{m},\cdots, h^{(N)}_{m} \right ]^M$,
$h^{(i)}_{m}=\alpha_{m,s}^{(i)}e^{-j\phi_{m,s}^{(i)}}$ denotes the channel coefficient of the $i^{th}$ element, $\alpha_{m,s}^{(i)}$ and $\phi_{m.s}^{(i)}$ denote the channel amplitude and phase, respectively, for the $\mathcal{S}-\mathcal{M}$ link.
Similarly, assuming that all the $\mathcal{M}-\mathcal{R}$ links are experienced Rician distribution, the channel vector is denoted by $\boldsymbol{g}_{m,s}$, where $\boldsymbol{g}_{m,s}$=$\left [ g^{(1)}_{m,s},\cdots, g^{(i)}_{m,s},\cdots, g^{(N)}_{m,s} \right ]^M$, $g^{(i)}_{m,s}=\beta_{m,s}^{(i)}e^{-j\Phi_{m,s}^{(i)}}$ is the channel coefficient of the $i^{th}$ element, $\beta_{m,s}^{(i)}$ and $\Phi_{m,s}^{(i)}$ defines the channel amplitude and phase, respectively, but for the $\mathcal{M}-\mathcal{R}$ link. 
In this case, maximizing the received SNR is achieved by adjusting the induced phase of RISs. This optimization involves the necessary phase cancelations and the precise alignment of the reflected signals originating from the RISs. Hence, the instantaneous SNR of the $\mathcal{S}-\mathcal{M}-\mathcal{R}$ link can be written as 
\begin{align}
    \gamma_{{s}}=\frac{{T_{x}}\left ( \sum_{i=1}^{N}\alpha_{m,s}^{(i)}\beta_{m,s}^{(i)} \right )^2}{N_{x}}=\bar{\gamma_{s}}\left ( \sum_{i=1}^{N}\alpha_{m,s}^{(i)}\beta_{m,s}^{(i)} \right )^2,
\end{align}
where $\bar{\gamma_{s}}=\frac{{T_{x}}}{N_{x}}
$ denotes the average SNR, $T_{x}$ is the transmitted power and $N_{x}$ denotes the noise power, respectively for the $\mathcal{S}-\mathcal{M}-\mathcal{R}$ link. After receiving the incoming signal, $\mathcal{R}$ transforms the signal into optical form and retransmits it to $\mathcal{D}$ across the FSO link. Hence, the instantaneous SNR, $\gamma_{d}$ due to the $\mathcal{R}-\mathcal{D}$ link is expressed as 
\begin{align}
\gamma_{\textit{d}}=\bar{\gamma}_{d} \left \| \vartheta_{d} \right \|^{2},
\end{align}
where $\bar{\gamma}_{d}$ defines the average SNR, $\vartheta_{d}$ defines the channel gain  for the $\mathcal{R}-\mathcal{D}$ link. Here, $\mathcal{R}$ is placed on the roof of the building and uses the FSO link to deliver data to $\mathcal{D}$, providing high-bandwidth and low-latency connectivity.

\noindent It is important to mention that when communication occurs between $\mathcal{S}$ to $\mathcal{D}$ through $\mathcal{R}$, there is a potential security concern: an unintended eavesdropper ($\mathcal{E}$), could attempt to intercept the communication channel via $\mathcal{S}-\mathcal{M}-\mathcal{E}$ link. Similar to the previous one, it is assumed that the eavesdropper link undergoes the Rician distribution. Hence, the channel vector is denoted by $\boldsymbol{g}_{m,e}$, where $\boldsymbol{g}_{m,e}$=$\left [ g^{(1)}_{m,e},\cdots, g^{(i)}_{m,e},\cdots, g^{(N)}_{m,e} \right ]^M$, $g^{(i)}_{m,e}=\beta_{m,e}^{(i)}e^{-j\Phi_{m,e}^{(i)}}$ is the channel coefficient of the $i^{th}$ element, $\beta_{m,e}^{(i)}$ and $\Phi_{m,e}^{(i)}$ defines the channel amplitude and phase, respectively but for the $\mathcal{M}-\mathcal{E}$ link. Therefore, the instantaneous SNR, $\gamma_{e}$ can be expressed as
\begin{align}
\gamma_{\textit{e}}=\frac{{T_{z}}\left ( \sum_{i=1}^{N}\alpha_{m,s}^{(i)}\beta_{m,e}^{(i)} \right )^2}{N_{z}}=\bar{\gamma}_{e}\left ( \sum_{i=1}^{N}\alpha_{m,s}^{(i)}\beta_{m,e}^{(i)} \right )^2,
\end{align}
where $\bar{\gamma}_{e}=\frac{{T_{z}}}{N_{z}}$, $\bar{\gamma}_{e}$ is the average SNR, ${T_{z}}$ and ${N_{z}}$ have the similar definition as described before but for the $\mathcal{S}-\mathcal{M}-\mathcal{E}$ link. Now, utilizing the variable gain AF relay, the received instantaneous SNR of the proposed RIS-aided network can be expressed as \cite [Eq.(9)] {rakib2024ris}
 \begin{align}
    \gamma_{eq}\cong min \left \{{ \gamma_{s},\gamma_{d}} \right \}
\end{align}

\subsection{PDF and CDF of RF Channel}
Assume all RF links are subject to a Rician distribution since this distribution helps to model the real-world signal environment with strong direct paths and multipath components. Hence, the PDF and CDF of $\gamma_{j}$ is expressed as \cite[Eq.~(12-13)]{10313311}
\begin{align}
 f_{\gamma_{j}}(\gamma)=\frac{\gamma^\frac{\textit{a}_{j}-1}{2}e^{-{\left(\frac{\sqrt{\gamma}}{\textit{b}_{j}\sqrt{\bar{\gamma_{j}}}
}\right)}}}{2\textit{b}_{j}^{^{\textit{a}_{j}+1}}\Gamma\left ( \textit{a}_{j}+1 \right )\bar{\gamma}_{j}\frac{\textit{a}_{j}+1}{2}},
 \end{align}
 \begin{align}
 F_{\gamma_{j}}(\gamma)=\frac{\gamma \left(\textit{a}_{j}+1,\frac{\sqrt{\gamma}}{\textit{b}_{j}\sqrt{\bar{\gamma_{j}}}}\right)}{\Gamma \left ( \textit{a}_{j}+1 \right )},
 \label{eq cdf}
 \end{align}
where $j \in (s,e)$, $\gamma$ $(\cdot,\cdot)$ is the lower incomplete gamma function \cite [Eq.~(8.350.1)] {gradshteyn1988tables}, and $\Gamma$ ($\cdot$) denotes the Gamma operator. The variables $a_{j}$ and $b_{j}$ are related to the mean and variance of a Rician random variable ($\mathcal{R}_{j}$) which can be written as 
\begin{align} \nonumber
\textit{a}_{j}=\frac{N\left (\mathbb{E}(\mathcal{R}_{j}) \right )^2}{Var\left (\mathcal{R}_{j} \right )}-1,
\end{align}
\begin{align} \nonumber
\textit{b}_{j}=\frac{Var(\mathcal{R}_{j})}{\mathbb{E}
\big(\mathcal{R}_{j})},
 \end{align}
Here, the mean and variance of $\mathcal{R}_{j}$ can be written, respectively, as
 \begin{align}
 \nonumber
\mathbb{E}(\mathcal{R}_{j})&=\frac{\pi e^{-0.5\left(k_{1,j}+k_{2,j}\right)}}{4\sqrt{\Omega_{1,j}\Omega_{2,j}}}\biggl[\left(k_{1,j}+1\right)I_{0}
(\frac{k_{1,j}}{2})
\\ \nonumber
&+k_{1,j}I_{1}
(\frac{k_{1,j}}{2})\biggl]\biggl
[(k_{2,j}+1)I_{0}(\frac{k_{2,j}}{2})+k_{2,j}I_{1}(\frac{k_{2,j}}{2})\biggl],
 \end{align}
\begin{align}
\nonumber
Var\left ( \mathcal{R}_{j} \right ) &=\frac{1}{16\Omega _{1,j}\Omega _{2,j}} \biggl[16 \left(k_{1,j}+1 \right)
 \left(k_{2,j}+1 \right)-\pi^2 \frac{e^{-k_{1,j}}}{e^{k_{2,j}}} 
\\
\nonumber
& \times\left \{ \left ( k_{1,j}+1 \right )
I_{0}
\left(\frac{k_{1,j}}{2} \right)
+k_{1,j}I_{1}
\left (\frac{k_{1,j}}{2} \right ) \right \}^2 
\\
\nonumber
& \times \left \{\left ( k_{2,j}+1 \right )I_{0}\left (\frac{k_{2,j}}{2} \right )
+k_{2,j}I_{1}\left (\frac{k_{2,j}}{2} \right ) \right \}^2 \biggl],
 \end{align}
 where $k_{1,j}$ is the shape parameter and and $\Omega_{1,j}$ is the scale parameter due to the  $\mathcal{S}-\mathcal{M}$ link. Similarly, $k_{2,j}$ is the shape parameter and and $\Omega _{2,j}$ is the scale  parameter for the  $\mathcal{M}-\mathcal{R}$ and  $\mathcal{M}-\mathcal{E}$ link, respectively, $\boldsymbol{I}_{\boldsymbol{v}}$ denotes the  modified $\boldsymbol{v}$ order first kind Bessel function \cite [Eq.~(8.431)]{gradshteyn1988tables}, and $\boldsymbol{k}_{\boldsymbol{v}}$ defines the modified $\boldsymbol{v}$ order second kind Bessel function \cite [Eq.~(8.432)]{gradshteyn1988tables}.

\subsection{PDF and CDF of FSO Link}
We assume the FSO link is subjected to Málaga turbulent distributions since this model is one of the most widely accepted fading models in FSO communication for exceptional characteristics. Therefore, the PDF of $\gamma_{d}$ is written as \cite[Eq.~(12)]{10058969}
\begin{align}
f_{\gamma _{d}}\left (\gamma \right )=\frac{X_{d}}{\gamma }\sum_{m_{d}=0}^{\beta_{d}}V_{{m}_{d}}G_{1,3}^{3,0}\left [ \bar{w}_{d}\left( \frac{\gamma}{U_{d}} \right)^\frac{1}{r_{d}}
\bigg|
\begin{array}{c}
\xi_{d}^2+1\\
\xi_{d}^2,\alpha_{d}, m_{d}\\
\end{array} \right ],
\end{align}
where
\begin{align}
    \nonumber
    X_{d}=\frac{2^{1-{r}_{d}}\alpha _{d}^\frac{\alpha _{d}}{2}\xi_{d}^2}{g_{d}^{1+\frac{\alpha_{d}}{2}}\Gamma(\alpha_{d})}\left ( \frac{g_{d}\beta _{d}}{g_{d}\beta _{d}+\mathcal{\omega} _{d}} \right )^{\beta _{d}+\frac{\alpha _{d}}{2}},
\end{align}
\begin{align}
\nonumber
\bar{w}_{d}=\frac{\xi_{d}^2\alpha_{d}\beta _{d}\left(g_{d}+\mathcal{\omega} _{d}\right)}{\left ( \xi_{d}^2+1 \right ) \left(g_{d}\beta_{d}+\mathcal{\omega} _{d}\right)},
    \end{align}
    \begin{align}
    \nonumber
 V_{{m}_{d}}=U_{{m}_{d}}
   \times \left(\frac{\alpha_{d}\beta_{d}}{g_{d}\beta_{d}+\mathcal{\omega}_{d}}\right)^{-\frac{\alpha_{d}+m_{d}}{2}},
    \end{align}
\begin{align}
\nonumber
U_{{m}_{d}}=\binom{\beta_{d}-1}{m_{d}-1}\frac{\left(g_{d}\beta_{d}+\mathcal{\omega}_{d}\right)^{1-\frac{m_{d}}{2}}}{\left(m_{d}-1\right )!}
\left(\frac{\mathcal{\omega}_{d}}{g_{d}}\right)^{m_{d}-1}\left(\frac{\alpha_{d}}{\beta_{d}}\right)^\frac{m_{d}}{2}.
\end{align}
Here, $(\alpha_{d}, \beta_{d})$ denotes the atmospheric turbulence,
$\xi_{d}$ denotes pointing error, $r_{d}$ represents the detection technique (i.e., $r_{d}=1$ defines the HD technique and $r_{d}=2$ defines the IM/DD technique), the electrical SNR is denoted by $U_d$ which is related to the average SNR of the FSO link, $\bar{\gamma}_{d}$ where ${U_{1}}=\bar{\gamma}_{d}$ (HD technique) and
${U_{2}}=\frac{\alpha_{d}\xi_{d}^2(\xi_{d}^2+1)^{-2}(\xi_{d}^2+2)(g_{d}+ \omega_{d})\bar{\gamma}_{d}}{(\alpha _{d}+1)[2g_{d}(g_{d}+2\omega _{d})+\omega _{d}^2(1+\frac{1}{\beta _{d}})]}$ (IM/DD technique), and $g_{d}$ is the average power received from off-axis eddies within the FSO connection. For the FSO link, ${\omega_{d}} = c_{d} + 2V_{d\zeta} + \sqrt{2V_{d\zeta}c_{d}}\cos(\theta_{x_{d}}-\theta_{y_{d}})$ denotes the coherent contributions in the average power, $c_{d}$= $2V_{d\zeta}(1-\zeta)$ characterizes the average power of the LOS component, $2V_{d}$ denotes scattered components of average power, $\zeta$ denotes the fraction of LOS-linked scattering power components within the range of 0 $\le \zeta \le 1$, and the deterministic loss phases is indicated by $\theta_{x_{d}}$ and $\theta_{y_{d}}$, $G_{.,.}^{.,.}[\cdot]$ represents the Meijer's G function as defined in \cite{gradshteyn1988tables}.
The  CDF of $\gamma_{d}$  is defined as 
\begin{align}
    F_{\gamma _{d}}(\gamma)=\sum_{m_{d}=0}^{\beta _{d}} Z_{d} W_{m_{d}}G_{r_{d}+1,3r_{d}+1}^{3r_{d},1}\left[\frac{h_{d}\gamma}{U_{d}}
\bigg|
\begin{array}{c}
1,L_{d1}\\
L_{d2},0\\
\end{array}\right],
\label{cdfm}
\end{align}
where
    $Z_{d}=\frac{X_{d}}{(2\pi)^{r_{d}}-1}$, $W_{{m}_{d}}=V_{{m}_{d}}r_{d}^{\alpha_{d}+m_{d}-1}$, $h_{d}=\frac{\bar{w}_{d}^{{r_{d}}}}{r_{d}^{2r_{d}}}$ and the series of $L_{d1}$ and $L_{d2}$ are denoted as $L_{d1}=\Delta({\frac{\xi_{d}^2+1}{r_{d}}, \frac{\xi_{d}^2+r_{d}}{r_{d}}})$ including $r_{d}$ terms and $L_{d2}=\Delta({\frac{\xi_{d}^2}{r_{d}}, \frac{\xi_{d}^2+r_{d}-1}{r_{d}}), \Delta(\frac{\alpha_{d}}{r_{d}}, \cdots, \frac{\alpha _{d}+r_{d}-1}{r_{d}}}), \Delta(\frac{m_{d}}{r_{d}}, \cdots, \frac{m{d}+r_{d}-1}{r_{d}})$ including $3r_{d}$ terms.

 \subsection{RIS Selection Strategy}
Although numerous advantages of employing multiple RIS in wireless communication systems, a practical approach involves selecting a single RIS from $M$ RISs to facilitate communication. This approach not only maintains the benefits of multiple RISs but also ensures cost-effective and efficient transmission. In this selection process, careful consideration is given to select the best RIS to maximize signal strength. Hence, the maximum end-to-end SNR of the selected RIS can be expressed as \cite [Eq.~9]{aldababsa2023multiple} 
 \begin{align}
 \gamma_{j^*}=\max_{M=1,\cdots,M}{\gamma_{j}}.
     \label{select ris}
 \end{align}
 Since RISs are passive in nature, it is imperative to emphasize that $\mathcal{S}$ must perform channel estimation for all available channels to ensure RIS offering the highest SNR. It is assumed that $\mathcal{S}$ possesses the necessary knowledge of the channel-state information to facilitate the realization of Eq. \eqref{select ris}. Considering the order statistic theory, the CDF of $\gamma_{j^*}$ can be simplified as
 \begin{align}
      F_{\gamma_{j^*}(\gamma)}=\prod_{m}^{M} F_{\gamma_{j}(\gamma)}
      \label{eq cdfs}
 \end{align}
Therefore, using \cite [Eq.~8.354.1]{gradshteyn1988tables} in \eqref{eq cdf} and
 substituting into \eqref{eq cdfs} 
and then applying multinomial theorem the  CDF of $\gamma_{j^*}$ can be expressed as
 \begin{align} \nonumber
    F_{\gamma^*_{j}(\gamma)}&=\sum_{\textit{k}_{0}+\textit{k}_{1}+ \cdots +\textit{k}_{\infty}=M}^{\infty}\binom{M}{\textit{k}_{0},\textit{k}_{1}, \cdots, \textit{k}_{\infty}}\prod_{n_{j}}^{} \left({c}_{n_{j}}\right)^{\textit{k}_{n_{j}}}
    \\
    & \times \gamma ^{MP_{j}+\frac{1}{2} \sum_{n_{j}}^{}n_{j}\textit{k}_{n_{j}}},
     \label{eq CDFs}
 \end{align}
 where $c_{n_{j}}$=$\frac{(-1)^{n_{j}}}{n_{j}!(\textit{a}_{j}+n_{j}+1)\Gamma (\textit{a}_{j}+1)}\left(\frac{1}{\textit{b}_{j}\sqrt{\bar{\gamma}}}\right)^{\textit{a}_{j}+n_{j}+1}$ and $P_{j}=\frac{\textit{a}_{j} +1}{2}$.  
 Finally, the PDF of $\gamma_{j^*}$ can be expressed as
\begin{align}
  f_{\gamma_{j^*}(\gamma)}=\frac{d}{d\gamma} F_{\gamma_{j^*}(\gamma)}=\frac{d}{d\gamma}\prod_{m}^{M} F_{\gamma_{j}(\gamma)}.
\end{align}
 Thus, the PDF of $\gamma_{j^*}$ is written finally as
 \begin{align}
 \nonumber
f_{\gamma_{j^*}(\gamma)}&=\sum_{\textit{k}_{0}+\textit{k}_{1}+\cdots+\textit{k}_{\infty}=M-1}^{\infty}\binom{M-1}{\textit{k}_{0},\textit{k}_{1}, .....\textit{k}_{\infty}} 
\\
 & \times\prod_{n_{j}}^{} ({c}_{n_{j}})^{\textit{k}_{n_{j}}} \delta_{j} \gamma ^{P_{j}(M-1))+\frac{1}{2} (\textit{a}_{j}-1+\sum_{n_{j}}^{}n_{j}\textit{k}_{n_{j}})},
    \label{pdfs}
 \end{align}
 where $\delta_j$=$\frac{Me^{-\frac{\sqrt{\gamma}}{\textit{b}_{j}\sqrt{\bar\gamma}}}}{2\textit{b}_{j}^{2{P_j}}\Gamma(2P_{j})\bar\gamma^{P_j}}$.

\subsection{CDF of SNR for Dual-Hop Model }
The CDF of ${{\gamma}_{eq}}$ can be defined as \cite[Eq.~(12)]{badrudduza2021security}
  \begin{align}
      F_{{\gamma}_{eq}}(\gamma)=F_{\gamma_{s^*}}(\gamma)+F_{\gamma_{d}}(\gamma)-F_{\gamma_{s^*}}(\gamma)F_{\gamma_{d}}(\gamma).
      \label{dual hop}
  \end{align}
 Substituting \eqref{eq CDFs} and \eqref{cdfm} into \eqref{dual hop}, the CDF of $\gamma_{eq}$ is expressed finally as
  \begin{align}
 \nonumber
     & F_{{\gamma}_{eq}}(\gamma) =\sum_{\textit{k}_{0}+\textit{k}_{1}+ \cdots +\textit{k}_{\infty}=M}^{\infty}\binom{M}{\textit{k}_{0},\textit{k}_{1}, \cdots, \textit{k}_{\infty}} \prod_{n_{s}}^{} \left({c}_{n_{s}}\right)^{\textit{k}_{n_{s}}}
      \\
      \nonumber
& \gamma^{MP_{s}+\frac{1}{2} \sum_{n_{s}}^{}n_{s}k_{n_{s}}}+\sum_{m_{d}=0}^{\beta _{d}}
 G_{r_{d}+1,3r_{d}+1}^{3r_{d},1}\left[\frac{h_{d}\gamma}{U_{d}}
\bigg|
\begin{array}{c}
1,L_{d1}\\
L_{d2},0\\
\end{array}\right]
\\
\nonumber
 & Z_{d}W_{m_{d}} -\sum_{\textit{k}_{0}+\textit{k}_{1}+ \cdots +\textit{k}_{\infty}=M}^{\infty}\sum_{m_{d}=0}^{\beta _{d}}  \binom{M}{\textit{k}_{0},\textit{k}_{1}, \cdots, \textit{k}_{\infty}} \prod_{n_{s}}^{} \left({c}_{n_{s}}\right)^{\textit{k}_{n_{s}}}
 \\
 & G_{r_{d}+1,3r_{d}+1}^{3r_{d},1}\left[\frac{h_{d}\gamma}{U_{d}}
\bigg|
\begin{array}{c}
1,L_{d1}\\
L_{d2},0\\
\end{array}\right] Z_{d}W_{m_{d}}
     \gamma ^{MP_{s}+\frac{1}{2} \sum_{n_{s}}^{}n_{s}\textit{k}_{n_{s}}}.
\label{Fso}
 \end{align}

\section{Performance Metrics} \label{pm}
In this section, the novel analytical expressions of various secrecy metrics such as SOP, ASC and EST are derived in terms of Meijer's $G$ function. To the authors' best knowledge, these derived expressions are novel and did not incorporate any existing literature. Furthermore, the asymptotic expressions at high SNR are also provided in this section.

\subsection{Lower Bound of SOP}
SOP measures the probability of unauthorized interception or eavesdropping on confidential information during transmitted data. When the secrecy capacity $(C_{sc})$ exceeds the target secrecy rate ($T_{Rs}$), it is assumed that the SOP is secured. Therefore, SOP can be defined mathematically as \cite [Eq.~36]{ibrahim2021enhancing}
\begin{align} 
    SOP_L=P_{r} \{\gamma_{eq} \leq \phi\gamma_{e}\}=\int_{0}^{\infty}F_{{\gamma}_{eq}}(\phi\gamma)f_{\gamma_{{e^*}} }(\gamma)d\gamma
    \label{SOP1},
\end{align} 
where $\phi$ = $2^{T_{Rs}}$. Substituting \eqref{Fso} and \eqref{pdfs} into \eqref{SOP1}, SOP is finally derived as (\ref{SOP}), $\mathcal{R}_{1}$ and $\mathcal{R}_{2}$ are the two integral terms that are derived as follows:
\begin{figure*}
    \begin{align} \label{SOP} \nonumber
   SOP_L&=\sum_{m_{d}=0}^{\beta _{d}}\sum_{\textit{k}_{0}+\textit{k}_{1}+\cdots+\textit{k}_{\infty}=M-1}^{\infty} \binom{M-1}{\textit{k}_{0},\textit{k}_{1}, \cdots, \textit{k}_{\infty}}\prod_{n_{e}}^{}\frac{Z_{d}W_{m_{d}}}{a^{-(q+1)}}{c}_{n_{e}}^{\textit{k}_{n_{e}}}\delta_{e}\bigg\{G_{3r_{d}+2,3r_{d}+2}^{3r_{d},2}
\left[\frac{h_{d}\phi}{U_{d}}a
\bigg|
\begin{array}{c}
-q,1,L_{d1}\\
L_{d2},0,-(q+1)\\
\end{array}\right]
\\ \nonumber
&+G_{r_{d}+2,3r_{d}+2}^{3r_{d}+1,1}
\left[\frac{h_{d}\phi}{U_{d}}a
\bigg|
\begin{array}{c}
1,L_{d1},-q\\
-(q+1),L_{d2},0\\
\end{array}\right]\bigg\}-\sum_{m_{d}=0}^{\beta _{d}}\sum_{\textit{k}_{0}+\textit{k}_{1}+ \cdots +\textit{k}_{\infty}=M}^{\infty}\sum_{\textit{k}_{0}+\textit{k}_{1}+\cdots+\textit{k}_{\infty}=M-1}^{\infty}\binom{M}{\textit{k}_{0},\textit{k}_{1}, \cdots, \textit{k}_{\infty}}
    \\
    \nonumber
    & \times\binom{M-1}{\textit{k}_{0},\textit{k}_{1}, \cdots, \textit{k}_{\infty}} 
\prod_{n_{s}}^{}\prod_{n_{e}}^{}\frac{Z_{d}W_{m_{d}}}{a^{-(c+1)}} {c}_{n_{s}}^{\textit{k}_{n_{s}}}{c}_{n_{e}}^{\textit{k}_{n_{e}}}\delta_{e}\phi^{MP_{s}+\frac{1}{2} \sum_{n_{s}}^{}n_{s}\textit{k}_{n_{s}}}\bigg\{
G_{3r_{d}+2,3r_{d}+2}^{3r_{d},2}
\Bigg[\frac{h_{d}\phi}{U_{d}}a
\bigg|
\begin{array}{c}
-c,1,L_{d1}\\
L_{d2},0,-(c+1)\\
\end{array}\Bigg]
\\
&+G_{r_{d}+2,3r_{d}+2}^{3r_{d}+1,1}
\left[\frac{h_{d}\phi}{U_{d}}a
\bigg|
\begin{array}{c}
1,L_{d1},-c\\
-(c+1),L_{d2},0\\
\end{array}\right]\bigg\},
\end{align}
\hrulefill
\end{figure*}



\subsubsection{Derivation of $\mathcal{R}_{1}$}
$\mathcal{R}_{1}$ is expressed as
\begin{align}
  \mathcal{R}_{1} =\int_{0}^{\infty}\gamma ^q G_{r_{d}+1,3r_{d}+1}^{3r_{d},1}\left[\frac{h_{d}\phi\gamma}{U_{d}}
\bigg|
\begin{array}{c}
1,L_{d1}\\
L_{d2},0\\
\end{array}\right] d\gamma
\label{r1},
\end{align}
where $q=P_{e}\left(M-1\right)+\frac{1}{2} \left(\textit{a}_{e}-1+\sum_{n_{e}}^{}n_{e}\textit{k}_{n_{e}}\right)$. With the aid of \cite[Eq.~(07.34.21.0084.01) and Eq.~(07.34.21.0085.01)] {meijerfunction}, $\mathcal{R}_{1}$ is obtained finally as
    \begin{align}
    \nonumber
   \mathcal{R}_{1} &=\frac{1}{a^{-(q+1)}}\bigg\{G_{3r_{d}+2,3r_{d}+2}^{3r_{d},2}
\left[\frac{h_{d}\phi}{U_{d}}a
\bigg|
\begin{array}{c}
-q,1,L_{d1}\\
L_{d2},0,-(q+1)\\
\end{array}\right]
\\
&+G_{r_{d}+2,3r_{d}+2}^{3r_{d}+1,1}
\left[\frac{h_{d}\phi}{U_{d}}a
\bigg|
\begin{array}{c}
1,L_{d1},-q\\
-(q+1),L_{d2},0\\
\end{array}\right]\bigg\}.
\label{r1+r2}
\end{align}
\subsubsection{Derivation of $\mathcal{R}_{2}$}

$\mathcal{R}_{2}$ is expressed as
\begin{align}
    \mathcal{R}_{2} =\int_{0}^{\infty}\gamma ^cG_{r_{d}+1,3r_{d}+1}^{3r_{d},1}\left[\frac{h_{d}\phi\gamma}{U_{d}}
\bigg|
\begin{array}{c}
1,L_{d1}\\
L_{d2},0\\
\end{array}\right] d\gamma,
\label{r_{2}}
\end{align}
where $c=MP_{s} +\frac{1}{2} \sum_{n_{s}}^{}n_{s}\textit{k}_{n_{s}}+{P_{e}\left(M-1\right)+\frac{1}{2} \left(\textit{a}_{e}-1+\sum_{n_{e}}^{}n_{e}\textit{k}_{n_{e}}\right)}.$
 According to the \cite[Eq.~ (07.34.21.0084.01) and Eq.~(07.34.21.0085.01)] {meijerfunction}, $\mathcal{R}_{2}$ is obtained finally as
    \begin{align}
    \nonumber
\mathcal{R}_{2} &=\frac{1}{a^{-(c+1)}}\bigg\{
G_{3r_{d}+2,3r_{d}+2}^{3r_{d},2}
\Bigg[\frac{h_{d}\phi}{U_{d}}a
\bigg|
\begin{array}{c}
-c,1,L_{d1}\\
L_{d2},0,-(c+1)\\
\end{array}\Bigg]
\\
&+G_{r_{d}+2,3r_{d}+2}^{3r_{d}+1,1}
\left[\frac{h_{d}\phi}{U_{d}}a
\bigg|
\begin{array}{c}
1,L_{d1},-c\\
-(c+1),L_{d2},0\\
\end{array}\right]\bigg\}.
\label{R2}
\end{align}

\subsubsection*{Asymptotic Expression}
Asymptotic expressions are used to understand the behavior of various performance metrics in high SNR regions. Therefore, the asymptotic expression for the lower bound of SOP is derived by applying \cite[Eq.~41]{ansari2015performance} to \eqref{SOP}. Finally, the asymptotic SOP can be expressed as shown in (\ref{sasy}), where
$P= \frac{h_{d}\phi a}{U_{d}}$, $S_{1}=\{-q, 1, L_{d_{1}}\}$, $S_{2}=\{L_{d_{2}}, 0, -(q+1)\}$, $S_{3}= \{1, L_{d_{1}}, -q\}$,  $S_{4}= \{-(q+1), L_{d_{2}}, 0\}$, $S_{5}=\{-c, 1, L_{d_{1}}\}$, $S_{6}=\{L_{d_{2}}, 0, -(c+1)\}$, $S_{7}= \{1, L_{d_{1}}, -c\}$, and $S_{8}= \{-(c+1), L_{d_{2}}, 0\}$ . 

\begin{figure*}
\begin{align} \label{sasy}
\nonumber
SOP_L({\infty})&=\sum_{m_{d}=0}^{\beta _{d}}\sum_{\textit{k}_{0}+\textit{k}_{1}+\cdots+\textit{k}_{\infty}=M-1}^{\infty} \binom{M-1}{\textit{k}_{0},\textit{k}_{1}, \cdots, \textit{k}_{\infty}}\prod_{n_{e}}^{}\frac{Z_{d}W_{m_{d}}}{a^{-(q+1)}}{c}_{n_{e}}^{\textit{k}_{n_{e}}}\delta_{e}
\bigg\{{\sum_{v_{1}=1}^{2}P^{S_{1},_{v_{1}}-1}}
   \times \frac{\prod_{l_{1}=1;l_{1}\neq v_{1}}^{2}\Gamma(S_{1,v_{1}}-S_{1,l_{1}})}{\prod_{l_{1}=3}^{3_{r_{d}}+2}\Gamma(1+S_{1,l_{1}}-S_{1,v_{1}})}
   \\
   \nonumber &\times\frac{\prod_{l_{1}=1}^{3_{r_{d}}}\Gamma(1+S_{2,l_{1}}-S_{1,v_{1}})}{{\prod_{l_{1}=3_{r_{d}}+1}^{3_{r_{d}}+2}\Gamma(S_{1,v_{1}}-S_{2,l_{2}})}}+{\sum_{v_{2}=1}^{1}P^{S_{3},_{v_{2}}-1}}
 \times\frac{\prod_{l_{2}=1;l_{2}\neq v_{2}}^{1}\Gamma(S_{3,v_{2}}-S_{3,l_{2}})\prod_{l_{2}=1}^{3_{r_{d}}+1}\Gamma(1+S_{4,l_{2}}-S_{3,v_{2}})}{\prod_{l_{2}=2}^{{r_{d}}+2}\Gamma(1+S_{3,l_{2}}-S_{3,v_{2}}){\prod_{l_{2}=3_{r_{d}}+2}^{3_{r_{d}}+2}\Gamma(S_{3,v_{2}}-S_{4,l_{2}})}} \bigg\}
 \\
 \nonumber
 &-\sum_{m_{d}=0}^{\beta _{d}}\sum_{\textit{k}_{0}+\textit{k}_{1}+ \cdots +\textit{k}_{\infty}=M}^{\infty}\sum_{\textit{k}_{0}+\textit{k}_{1}+\cdots+\textit{k}_{\infty}=M-1}^{\infty}\binom{M}{\textit{k}_{0},\textit{k}_{1}, \cdots,\textit{k}_{\infty}}
    \binom{M-1}{\textit{k}_{0},\textit{k}_{1},\cdots,\textit{k}_{\infty}} 
    \prod_{n_{s}}^{}\prod_{n_{e}}^{}\frac{Z_{d}W_{m_{d}}}{a^{-(c+1)}} 
{c}_{n_{s}}^{\textit{k}_{n_{s}}}{c}_{n_{e}}^{\textit{k}_{n_{e}}}\delta_{e}
\\
\nonumber
& \times \phi^{MP_{s}+\frac{1}{2} \sum_{n_{s}}^{}n_{s}\textit{k}_{n_{s}}}
\bigg\{{\sum_{v_{1}=1}^{2}P^{S_{5},_{v_{1}}-1}}
\frac{\prod_{l_{1}=1;l_{1}\neq v_{1}}^{2}\Gamma(S_{5,v_{1}}-S_{5,l_{1}})\prod_{l_{1}=1}^{3_{r_{d}}}\Gamma(1+S_{6,l_{1}}-S_{5,v_{1}})}{\prod_{l_{1}=3}^{3_{r_{d}}+2}\Gamma(1+S_{5,l_{1}}-S_{5,v_{1}}){\prod_{l_{1}=3_{r_{d}}+1}^{3_{r_{d}}+2}\Gamma(S_{5,v_{1}}-S_{6,l_{2}})}} +{\sum_{v_{2}=1}^{1}P^{S_{7},_{v_{2}}-1}}
   \\
& \times\frac{\prod_{l_{2}=1;l_{2}\neq v_{2}}^{1}\Gamma(S_{7,v_{2}}-S_{7,l_{2}})\prod_{l_{2}=1}^{3_{r_{d}}+1}\Gamma(1+S_{8,l_{2}}-S_{7,v_{2}})}{\prod_{l_{2}=2}^{{r_{d}}+2}\Gamma(1+S_{7,l_{2}}-S_{7,v_{2}}){\prod_{l_{2}=3_{r_{d}}+2}^{3_{r_{d}}+2}\Gamma(S_{7,v_{2}}-S_{8,l_{2}})}} \bigg\},
\end{align}
\hrulefill
\end{figure*}

\subsection{ASC Analysis} 
ASC is a measure of the average quantity of secured information to keep data confidential from eavesdroppers. It considers the characteristics of both authorized and unauthorized communication channels. Mathematically, it can be denoted as \cite [Eq.~19]{ibrahim2021enhancing}
\begin{align}
    ASC= \int_{0}^{\infty}\frac{F_{\gamma_{e^*}}(\gamma)}{1+\gamma }\left \{ 1- F_{{\gamma}_{eq}}(\gamma) \right \} d\gamma.
    \label{ASC}
\end{align}
Substituting \eqref{eq CDFs} and \eqref{Fso} into \eqref{ASC}, ASC is derived finally as (\ref{asc formula}), where the integral terms of $\mathcal{X}_{1}$, $\mathcal{X}_{2}$, $\mathcal{X}_{3}$, and $\mathcal{X}_{4}$ are derived as follows:
\begin{figure*}
    \begin{align} \nonumber
    ASC&=\sum_{\textit{k}_{0}+\textit{k}_{1}+ \cdots +\textit{k}_{\infty}=M}^{\infty}\binom{M}{\textit{k}_{0},\textit{k}_{1}, \cdots, \textit{k}_{\infty}}\prod_{n_{e}}^{} {c}_{n_{e}}^{\textit{k}_{n_{e}}}\biggl \{\frac{1}{a^{-(\mu_{e} +1)}}\bigg(G_{2,2}^{1,2}\left[a
\bigg|
\begin{array}{c}
-\mu_{e},0\\
0,-(\mu_{e}+1)\\
\end{array}\right]+G_{2,2}^{2,1}\left[a
\bigg|
\begin{array}{c}
0,-\mu_{e}\\
-(\mu_{e}  +1),0\\
\end{array}\right]\bigg)
\\ \nonumber
&-\sum_{\textit{k}_{0}+\textit{k}_{1}+ \cdots +\textit{k}_{\infty}=M}^{\infty}\binom{M}{\textit{k}_{0},\textit{k}_{1}, \cdots, \textit{k}_{\infty}}\prod_{n_{s}}^{} {c}_{n_{s}}^{\textit{k}_{n_{s}}}\frac{1 }{a^{-(\mu_{e}+\mu_{s} +1)}}\bigg(G_{2,2}^{1,2}\Bigg[a
\bigg|
\begin{array}{c}
-(\mu_{e}+\mu_{s}),0\\
0,-(\mu_{e}+\mu_{s}+1)\\
\end{array}\Bigg] G_{2,2}^{2,1}\left[a
\bigg|
\begin{array}{c}
0,-(\mu_{e}+\mu_{s})\\
-(\mu_{e}+\mu_{s}+1),0\\
\end{array}\right]\bigg)
\\ \nonumber
&-\sum_{m_{d}=0}^{\beta _{d}}Z_{d}W_{m_{d}}G_{r_{d}+2,3r_{d}+2}^{3r_{d}+1,2}\left[\frac{h_{d}}{U_{d}}
\bigg|
\begin{array}{c}
1,-\mu_{e},L_{d1}\\
L_{d2},-\mu_{e},0\\
\end{array}\right]+\sum_{m_{d}=0}^{\beta _{d}}Z_{d}W_{m_{d}}\sum_{\textit{k}_{0}+\textit{k}_{1}+ \cdots +\textit{k}_{\infty}=M}^{\infty}\binom{M}{\textit{k}_{0},\textit{k}_{1}, \cdots, \textit{k}_{\infty}}\prod_{n_{s}}^{} {c}_{n_{s}}^{\textit{k}_{n_{s}}}
\\ 
& G_{r_{d}+2,3r_{d}+2}^{3r_{d}+1,2}\left[\frac{h_{d}}{U_{d}}
\bigg|
\begin{array}{c}
1,-(\mu_{e}+\mu{s}),L_{d1}\\
L_{d2},-(\mu_{e}+\mu{s}),0\\
\end{array}\right] \biggl \},
    \label{asc formula}
\end{align}
\hrulefill
\end{figure*}


\subsubsection{Derivation of $\mathcal{X}_{1}$}
Utilizing the identities of \cite[Eq.~8.4.2.5]{art3} to transform $\frac{1}{1+\gamma}$ into Meijer's $G$, $\mathcal{X}_{1}$ is written as
\begin{align}
\nonumber
    \mathcal{X}_{1} &=\int_{0}^{\infty}\frac{1}{1+\gamma }\gamma ^{\mu_{e}}d\gamma &
    \\
    &=\int_{0}^{\infty }\gamma ^{\mu_{e}}G_{1,1}^{1,1}\left[\gamma
\bigg|
\begin{array}{c}
0\\
0\\
\end{array}\right] d\gamma,
\label{meijerG}
\end{align}
where  $\mu_{e}={MP_{e}+\frac{1}{2} \sum_{n_{e}}^{}n_{e}\textit{k}_{n_{e}}}$. Now, with the help of \cite[Eq.~(07.34.21.0084.01) and Eq.~07.34.21.0085.01] {meijerfunction}, $\mathcal{X}_{1}$ is finally derived as
\begin{align}
\nonumber
   \mathcal{X}_{1}&=\frac{1}{a^{-(\mu_{e} +1)}}\bigg\{G_{2,2}^{1,2}\left[a
\bigg|
\begin{array}{c}
-\mu_{e},0\\
0,-(\mu_{e}+1)\\
\end{array}\right]
\\
&+G_{2,2}^{2,1}\left[a
\bigg|
\begin{array}{c}
0,-\mu_{e}\\
-(\mu_{e}  +1),0\\
\end{array}\right]\bigg\}.
\label{X1}
\end{align}

\subsubsection{Derivation of $\mathcal{X}_{2}$}
Using the same procedure of $\mathcal{X}_1$, $\mathcal{X}_2$ is obtained as
\begin{align}
\nonumber
      \mathcal{X}_{2}&=\int_{0}^{\infty}\frac{1}{1+\gamma }\gamma ^{\mu_{e}+\mu_{s}}d\gamma 
     \\
         &=\int_{0}^{\infty }\gamma ^{\mu_{e}+\mu_{s}}G_{1,1}^{1,1}\left[\gamma
\bigg|
\begin{array}{c}
0\\
0\\
\end{array}\right] d\gamma,
\label{X2}
\end{align}
where  $\mu_{s}={MP_{s}+\frac{1}{2} \sum_{n_{s}}^{}n_{s}\textit{k}_{n_{s}}}$. Now, utilizing the similar identities as used in \eqref{X1}, $\mathcal{X}_2$ is finally expressed as
\begin{align}
\nonumber
  \mathcal{X}_{2}&=\frac{1 }{a^{-(\mu_{e}+\mu_{s} +1)}}\bigg\{G_{2,2}^{1,2}\Bigg[a
\bigg|
\begin{array}{c}
-(\mu_{e}+\mu_{s}),0\\
0,-(\mu_{e}+\mu_{s}+1)\\
\end{array}\Bigg]
\\
&+G_{2,2}^{2,1}\left[a
\bigg|
\begin{array}{c}
0,-(\mu_{e}+\mu_{s})\\
-(\mu_{e}+\mu_{s}+1),0\\
\end{array}\right]\bigg\}.
\end{align}

\subsubsection{Derivation of $\mathcal{X}_{3}$}
Similar to $\mathcal{X}_2$, $\mathcal{X}_{3}$ is expressed as
\begin{align}
\nonumber
      \mathcal{X}_{3}&=\int_{0}^{\infty}\frac{1}{1+\gamma }\gamma ^{\mu_{e}}G_{r_{d}+1,3r_{d}+1}^{3r_{d},1}\left[\frac{h_{d}\gamma}{U_{d}}
\bigg|
\begin{array}{c}
1,L_{d1}\\
L_{d2},0\\
\end{array}\right] d\gamma 
     \\
        & =\int_{0}^{\infty}\gamma ^{\mu_{e}}G_{1,1}^{1,1}\left[\gamma
\bigg|
\begin{array}{c}
0\\
0\\
\end{array}\right]G_{r_{d}+1,3r_{d}+1}^{3r_{d},1}\left[\frac{h_{d}\gamma}{U_{d}}
\bigg|
\begin{array}{c}
1,L_{d1}\\
L_{d2},0\\
\end{array}\right] d\gamma.
\label{x3}
\end{align}
Now, utilizing the identity of \cite [Eq.~07.34.21.0011.01]{meijerfunction}, $\mathcal{X}_3$ can be written finally as
\begin{align}
X_{3}=G_{r_{d}+2,3r_{d}+2}^{3r_{d}+1,2}\left[\frac{h_{d}}{U_{d}}
\bigg|
\begin{array}{c}
1,-\mu_{e},L_{d1}\\
L_{d2},-\mu_{e},0\\
\end{array}\right].
\end{align}

\subsubsection{Derivation of $\mathcal{X}_{4}$}
$\mathcal{X}_{4}$ is expressed as
\begin{align}
\nonumber
      \mathcal{X}_{4}=\int_{0}^{\infty }\gamma ^{\mu_{e}+\mu_{s}}G_{1,1}^{1,1}\left[\gamma
\bigg|
\begin{array}{c}
0\\
0\\
\end{array}\right]
\\
\times G_{r_{d}+1,3r_{d}+1}^{3r_{d},1}\left[\frac{h_{d}\gamma}{U_{d}}
\bigg|
\begin{array}{c}
1,L_{d1}\\
L_{d2},0\\
\end{array}\right] d\gamma .
\label{x4}
\end{align}
With the help of the similar identities as utilized in $\mathcal{X}_{3}$, $\mathcal{X}_{4}$ is finally expressed as
\begin{align}
\mathcal{X}_{4}=G_{r_{d}+2,3r_{d}+2}^{3r_{d}+1,2}\left[\frac{h_{d}}{U_{d}}
\bigg|
\begin{array}{c}
1,-(\mu_{e}+\mu{s}),L_{d1}\\
L_{d2},-(\mu_{e}+\mu{s}),0\\
\end{array}\right].
\end{align}
\subsection{EST Analysis} 
EST analysis in wireless communication is a measure used to evaluate the performance of secure wireless communication systems, particularly in the context of physical layer security. EST combines the concepts of secrecy rate and effective capacity to provide a more comprehensive measure of secure communication performance. Mathematically, it can be defined as \cite [Eq.~60]{10313311}
\begin{align}
    EST=T_{Rs}\left(1-SOP\right).
    \label{EST}
\end{align}
Now, substituting Eq. (\ref{SOP}) into  Eq. \eqref{EST}, the analytical expression of EST can be obtained easily.

\section{Numerical Results} \label{nr}
The objective of this section is to demonstrate the numerical results associated with the derived performance metrics (e.g., SOP, ASC, and EST) in order to investigate the impacts of various system parameters on the secrecy performance. Furthermore, the validation of theoretical expressions is conducted using Monte Carlo simulations that average $10^6$ random samples of Rician and Málaga random variables. The simulation results are in good agreement with the analytical results, indicating that our derived expressions are accurate. According to \cite{10313311,10058969}, the simulation parameters are set to: $k_{1,s}=k_{2,s}=k_{1,e}=k_{2,e}=2$, $\Omega_{1,s}=\Omega_{2,s}=\Omega_{1,e}=\Omega_{2,e}=1$, $(\alpha_{d}, \beta_{d})=(2.296, 2)$, $\xi_{d}=(1.1, 6.7)$, $r_{d}=(1, 2)$, $N=2$, $M=2$, $\bar{\gamma_{s}}=20$ dB, $\bar{\gamma_{d}}=25$ dB, $\bar{\gamma_{e}}=0$ dB, and $T_{Rs}=0.5$ bits/sec/Hz, unless specified otherwise. Note that  $(\alpha_{d}, \beta_{d})=(2.296,2)$ represents the strong turbulence, $(\alpha_{d},\beta_{d})=(4.2,3)$ represents moderate turbulence, and  $(\alpha_{d},\beta_{d})=(8,4)$ represents weak turbulence \cite{10058969}. Furthermore, an asymptotic analysis is performed, which shows close agreement with the analytical results in the high SNR regimes.   
\begin{figure}[!ht]
\centerline{\includegraphics[width=0.35\textwidth,angle =0]{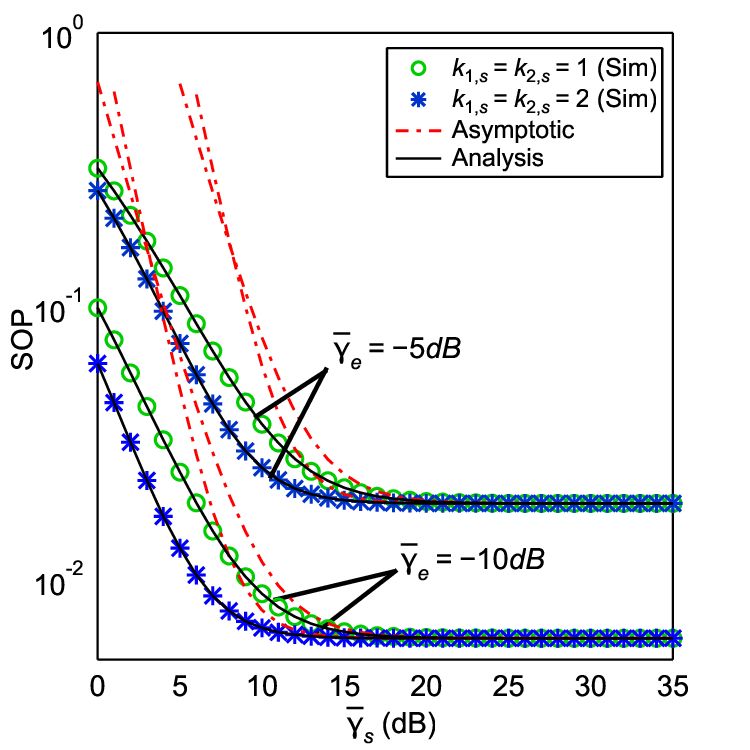}}
    \caption{SOP versus $ {\bar{\gamma}_{s}}$ for selected values of $K_{1, s}$=$K_{2, s}$.
    }
    \label{sop-rfs}
\end{figure}

\subsection{Impact of RF Fading Parameters}
\begin{figure}[!ht]
\centerline{\includegraphics[width=0.35\textwidth,angle =0]{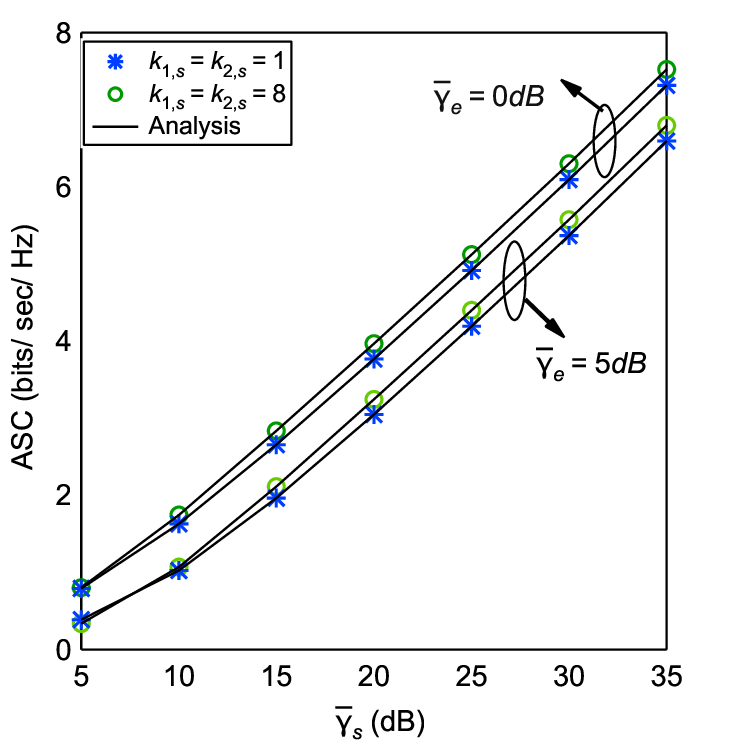}}
    \caption{ASC versus $\bar{\gamma}_{s}$
     for selected values of $k_{1,s}=k_{2,s}$.}
    \label{asc-rf1}
\end{figure}

Figs. \ref{sop-rfs} - \ref{asc-rf1} illustrate the impact of the shape parameter, $K$ on the $\mathcal{S}-\mathcal{M}-\mathcal{R}$ link. To analyze this effect, SOP and ASC are plotted against $\bar{\gamma_{s}}$ for different values of  $\bar{\gamma_{e}}$. The figures reveal that the SOP values decrease as 
$K_{1,s}$ and $K_{2,s}$ increase, while ASC values show an upward trend with increasing $K_{1,s}$ and $K_{2,s}$. This behavior is expected since a higher
$K$ value implies a stronger LOS component relative to scattered components, which diminishes the effects of random scattering and fading on the received signal. Consequently, the received signal power becomes more concentrated, enhancing signal quality at the relay, $\mathcal{R}$ and significantly improving the secrecy performance of the proposed model. As shown in the figure that the secrecy performance improves when $\bar{\gamma}_{s}$ is increased because a higher SNR indicates a stronger and more reliable signal at the legitimate receiver. This increased signal strength makes it more difficult for the eavesdropper to intercept and decode the transmitted information effectively, as the legitimate receiver can better distinguish the signal from noise and interference.

\begin{figure}[!ht]
\centerline{\includegraphics[width=0.35\textwidth,angle =0]{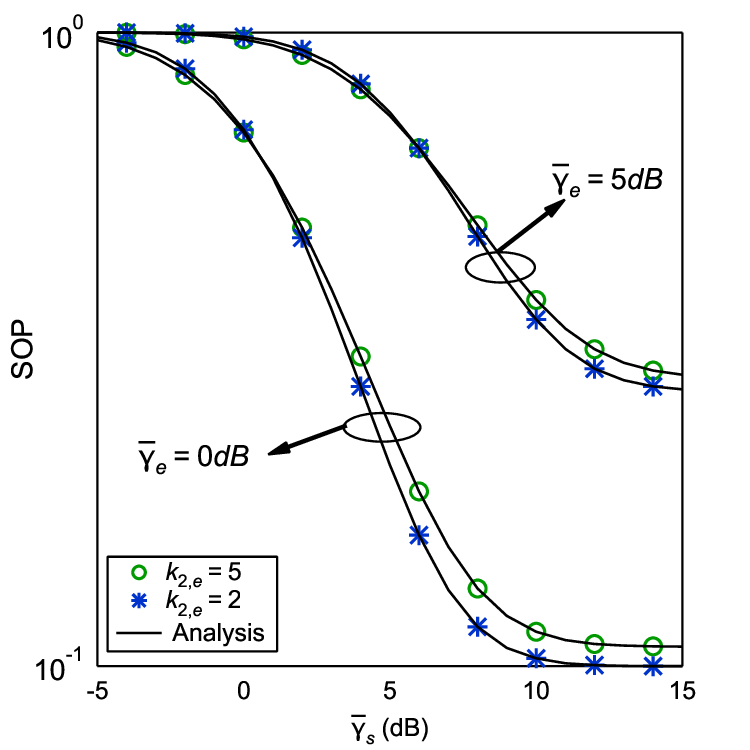}}
    \caption{SOP versus $\bar{\gamma_{s}}$ for selected values of $k_{2,e}$.
    }
    \label{sop-rf1}
\end{figure}

On the other hand, Fig. \ref{sop-rf1} depicts the impact of shape parameter, $K$ on the $\mathcal{S}-\mathcal{M}-\mathcal{E}$ link. Notably, SOP performance deteriorates significantly as $K_{2,e}$ increases from $2$ to $5$. This decline is attributed to a strengthening LOS component relative to scattered components in the eavesdropper channel. A dominant LOS component results in a more reliable eavesdropper signal with reduced fading, thereby enhancing eavesdropping capabilities and diminishing overall secrecy capacity. It is also observed that the outage probability increases significantly as $\bar{\gamma}_{e}$ rises from $0$ dB to $5$ dB. This is expected, as the higher 
$\bar{\gamma}_{e}$ enhances the strength of the $\mathcal{S}-\mathcal{M}-\mathcal{E}$ link, making it easier for the $\mathcal{E}$ to intercept confidential information. Hence, the secrecy performance degrades substantially.

\begin{figure}[!ht]
\centerline{\includegraphics[width=0.35\textwidth,angle =0]{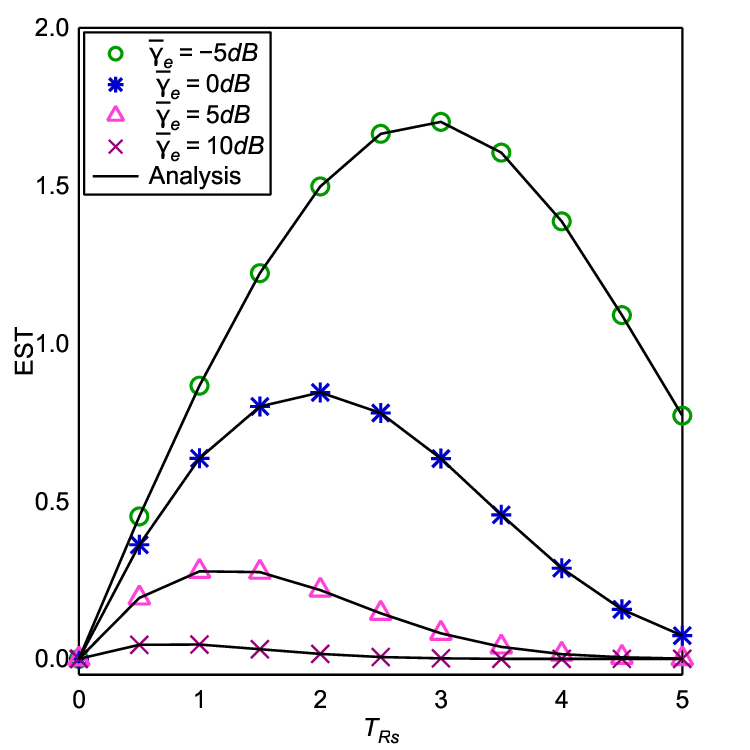}}
    \caption{EST versus ${T _{Rs}}$ for selected values of $\bar{\gamma}_{e}$.}
    \label{est}
\end{figure}
In Fig. \ref{est}, EST is plotted against $T_{Rs}$ to investigate the impact of secrecy rate under different values of $\bar{\gamma}_{e}$. It is observed that when plotting EST against the $T_{Rs}$, the resulting curve typically begins as a monotonic increasing function, reflecting the ability to maintain secure communication while increasing $T_{Rs}$. Initially, with an increase in the $T_{Rs}$, EST rises as well, indicating the capacity to sustain higher secure throughput. However, this upward trend continues only to a certain point. As $T_{Rs}$ approaches the capacity of the legitimate channel, the EST curve starts to flatten and eventually reaches a saturation point. Beyond this point, further increases in $T_{Rs}$ do not lead to significant gains in EST and cause a decrease when $T_{Rs}$ exceeds the ability to maintain security. This behavior occurs due to the inherent trade-offs in wireless communication systems when balancing secrecy and throughput. The legitimate communication channel has a finite capacity, and as the target secrecy rate increases, it demands more resources to maintain secure communication. Initially, the system can meet this demand, leading to a rise in EST. However, as the target rate approaches the capacity of the channel, the system struggles to maintain both high secrecy and throughput, causing the EST to flatten out. Eavesdroppers further complicate this, as the system must allocate more resources to secure communication against intercept attempts. As these resources are stretched thin, less is available to sustain high throughput, leading to saturation in EST. The system must optimize between security and performance, and when the target secrecy rate is too high, throughput may be sacrificed to maintain security, resulting in the observed decline in EST.

\begin{figure}[!ht]
\centerline{\includegraphics[width=0.35\textwidth,angle =0]{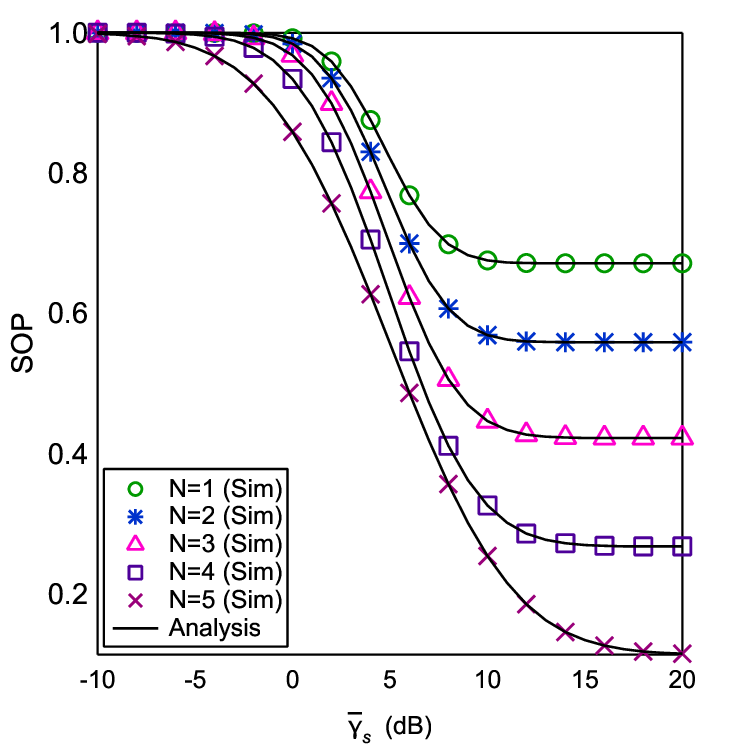}}
    \caption{SOP versus $\bar{\gamma}_{s}$ for selected values of $N$.
    }
    \label{r elements}
\end{figure}

\subsection{Impact of RIS}

\begin{figure}[!ht]
\centerline{\includegraphics[width=0.35\textwidth,angle =0]{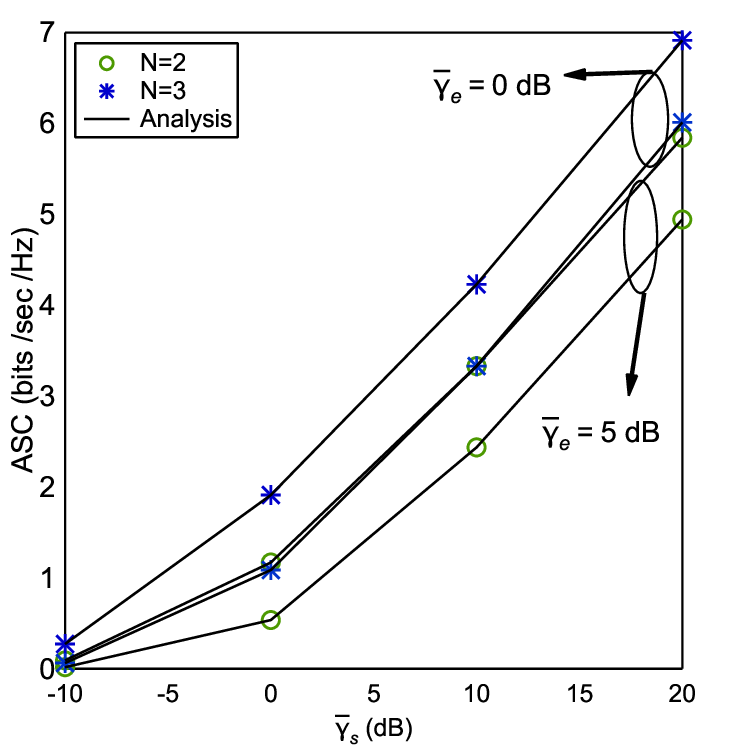}}
    \caption{ASC versus $\bar{\gamma}_{s}$ for selected values of $N$.
    }
    \label{asc ref}
\end{figure}

To analyze the impact of RIS elements, SOP and ASC is plotted against $\bar{\gamma}_{s}$ in Figs. \ref{r elements} - \ref{asc ref} due to the $\mathcal{S}-${M}$-\mathcal{R}$ link. It is clearly shown in Fig. \ref{r elements} that the value of SOP decreases when $N$ increases from $1$ to $5$. On the other hand, the value of ASC is improved with the increase of $N$. This is expected because with more reflecting elements, the RIS can precisely manipulate the reflected signals, directing them more effectively towards the relay, $\mathcal{R}$ while reducing their strength in the direction of potential eavesdropper, $\mathcal{E}$. This targeted beam-forming increases the channel gain between the source and the realy, resulting in a stronger and more reliable communication link. Additionally, the RIS can help to create destructive interference in the direction of eavesdroppers, further weakening their ability to intercept the signal. These factors combined lead to an improvement in the overall secrecy performance as the number of RIS elements increases. In other word, more reflecting elements can provide diversity in the received signal, which overcomes fading and improves reliability.

\begin{figure}[!ht]
\centerline{\includegraphics[width=0.35\textwidth,angle =0]{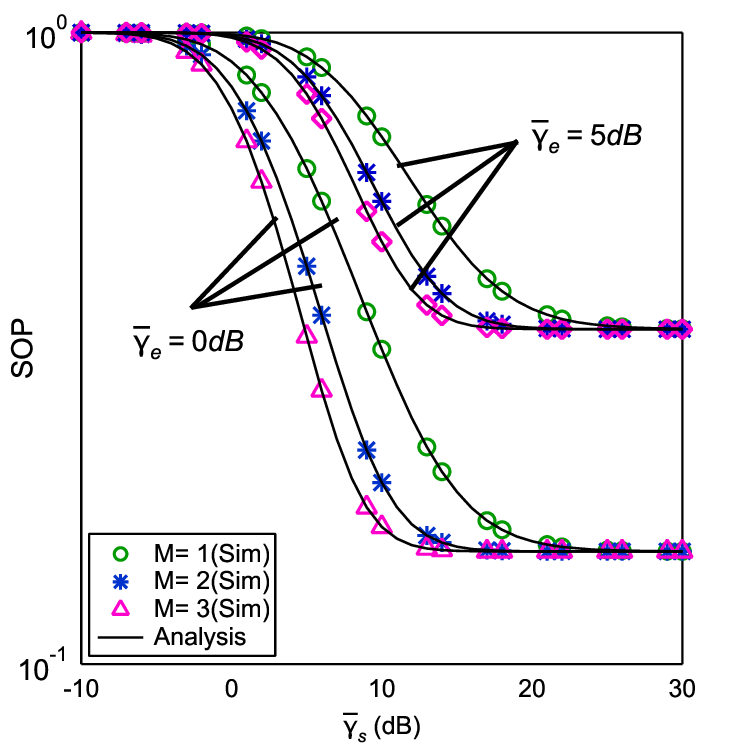}}
     \caption{SOP versus $\bar{\gamma}_{s}$ for selected values of $M$.
     }
   \label{multi ris1}
 \end{figure}

In Fig. \ref{multi ris1}, the impact of multiple RIS units, $M$ due to  the proposed model is investigated. For this purpose, SOP is plotted against $\bar{\gamma}_{s}$ under different values of $\bar{\gamma}_{e}$. Our findings indicate that SOP performance improves as the number of $M$ increases. This is anticipated because, in the proposed system, where a single RIS is used to transmit the signal to the receiver in the presence of an eavesdropper, increasing the available RIS units can enhance secrecy performance. With more RIS units to choose from, the system has a higher probability of selecting an RIS that not only provides optimal channel conditions for $\mathcal{R}$ but also minimizes exposure to $\mathcal{E}$. This selection process enhances the ability to direct the signal towards $\mathcal{R}$ while reducing its strength or creating destructive interference in the direction of $\mathcal{E}$. As a result, the signal received by $\mathcal{E}$ becomes weaker, reducing its ability to intercept or decode the transmitted information. Moreover, the increased number of RIS units introduces greater spatial diversity, enabling the system to more effectively counter eavesdropping attempts by exploiting the most favorable transmission paths. This enhanced ability to select the most secure RIS results in a substantial improvement in secrecy performance, reducing the chances of successful interception and consequently ensuring higher secrecy rates.

\begin{figure}[!ht]
\centerline{\includegraphics[width=0.35\textwidth,angle =0]{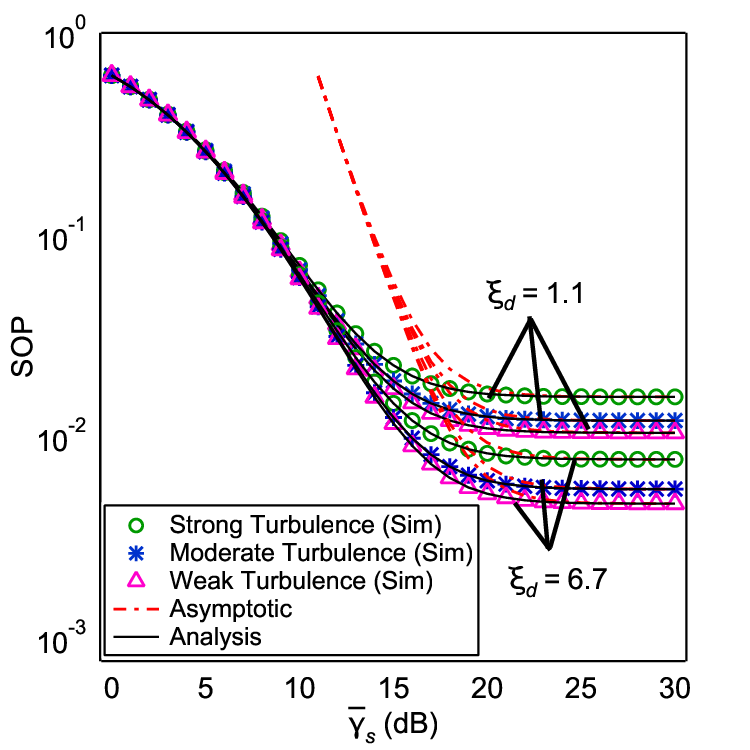}}
    \caption{SOP versus $\bar{\gamma}_{s}$ for selected values of $(\alpha_{d},\beta_{d})$.
    }
    \label{turbulence}
\end{figure}
\subsection{Impact of FSO Parameters }

\begin{figure}[!ht]
\centerline{\includegraphics[width=0.35\textwidth,angle =0]{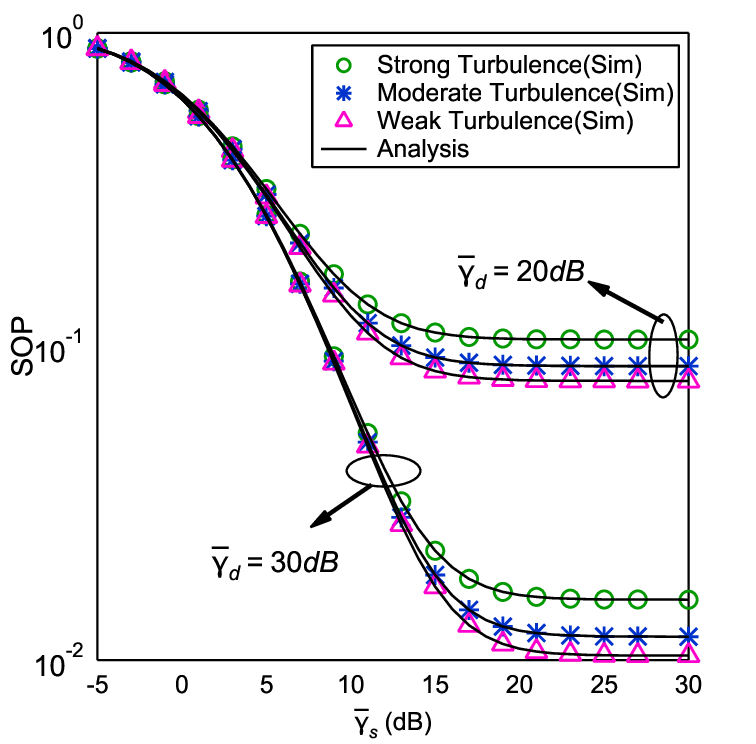}} \caption{SOP versus $\bar\gamma_{s}$ for selected values of ($\alpha_{d}$,$\beta_{d}$).}
    \label{fso tur}
 \end{figure}

The impact of FSO turbulence conditions on the $\mathcal{R}-\mathcal{D}$ link is demonstrated in Figs. \ref{turbulence} - \ref{fso tur}. Both figures demonstrate that increasing the values of $(\alpha_{d}, \beta_{d})$ mitigates the effects of atmospheric turbulence, leading to improved SOP performance. This is anticipated because, as turbulence diminishes, the reliability and predictability of the communication channel increase.  In the presence of strong turbulence, the transmitted signal undergoes severe distortions, resulting in significant errors and inconsistencies in the received signal. Conversely, under weak turbulence conditions, the signal exhibits greater stability and consistency, enabling improved synchronization between the relay and receiver. Furthermore, the reduced signal fluctuations in weak turbulence conditions hinder the ability of the eavesdropper to accurately estimate the legitimate channel, thereby enhancing the capacity of the system to maintain a secure communication link.

To evaluate the impact of pointing error, $\xi_{d}$ on the $\mathcal{R}-\mathcal{D}$ link, we plotted the SOP against the $\bar{\gamma}_{s}$ in Fig. \ref{turbulence}. Our findings demonstrate that SOP decreases as $\xi_{d}$ increases from $1.1$ to $6.7$. This suggests that increasing $\xi_{d}$ effectively mitigates the impact of pointing error at $\mathcal{D}$ by improving the accuracy and stability of signal alignment between the relay, $\mathcal{R}$ and the receiver, $\mathcal{D}$. Pointing errors, which occur when the optical beam misaligned with the receiver aperture, result in significant power losses. This degradation in signal strength compromises communication quality. As a result, the system becomes more vulnerable to interception. Minimizing pointing errors ensures that the optical beam is more accurately directed towards $\mathcal{D}$, resulting in a stronger and more consistent signal. Consequently, the secrecy performance of the proposed model is significantly enhanced.

Fig. \ref{fso tur} illustrates the influence of $\bar{\gamma}_{d}$ on secrecy performance. The figure reveals a substantial decrease in outage probability when $\bar{\gamma}_{d}$ increases from $20$ dB to $30$ dB. This reduction is attributed to the higher SNR, indicating a stronger, clearer signal with reduced noise interference at the receiver, $\mathcal{D}$.  Increasing $\bar{\gamma}_{d}$ strengthens the transmitted signal from the relay, thereby enhancing the overall signal strength of the communication. This improved signal quality hinders the ability of the eavesdropper to intercept and decode the information accurately. Consequently, the secrecy performance of the proposed model is significantly enhanced.

\begin{figure}[!ht]
\centerline{\includegraphics[width=0.35\textwidth,angle =0]{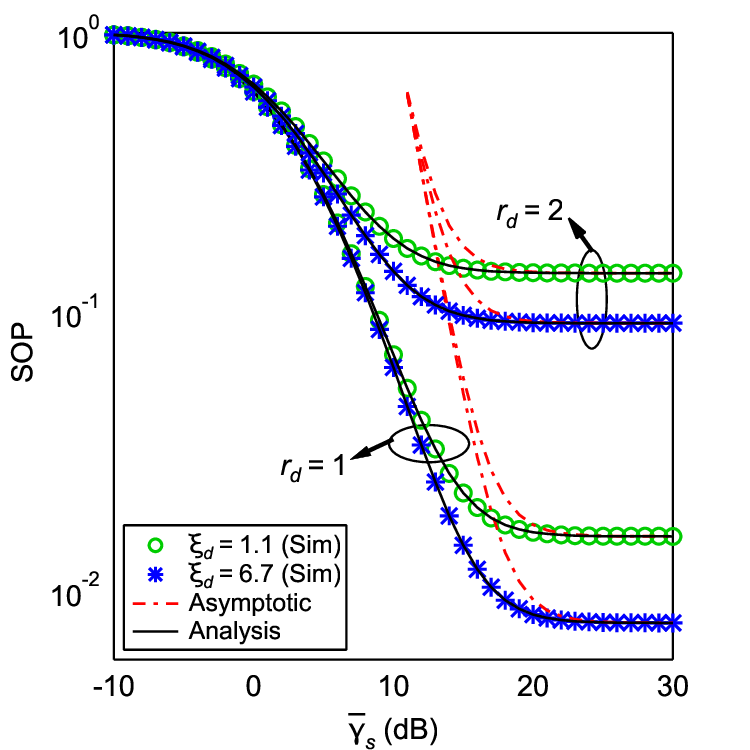}}
    \caption{SOP versus $\bar\gamma_{s}$ for selected values of $\xi_{d}$ and $r_d$.
    }
    \label{det}
\end{figure}

To evaluate the influence of receiver detection techniques on system performance, we plotted the SOP against $\bar\gamma_{s}$ in Fig. \ref{det}. Our findings demonstrate that the HD technique outperforms IM/DD in terms of secrecy performance. Additionally, HD technique enables more accurate signal reconstruction and a higher SNR.  This enhanced detection capability empowers the legitimate receiver, $\mathcal{D}$, implemented with the HD technique, to decode the transmitted signal with greater accuracy, even in the face of noise or interference. Conversely, IM/DD relies on the intensity of the received signal, making it more vulnerable to noise, turbulence, and other channel impairments. This susceptibility results in a lower SNR and increased vulnerability to eavesdropping. Consequently, the increased sensitivity and robustness of HD contribute significantly to its superior secrecy performance within the proposed model.

\subsection{Design Guidelines}
Based on the findings from the investigation into the secrecy performance of the proposed system, the following design guidelines should be considered:
\begin{itemize}
    \item In Fig. \ref{r elements}, with $N=1$, the SOP is $0.6755$. When the RIS elements increase to $N=5$ at an SNR of $10$ dB, the SOP decreases to $0.2543$. This represents a $63\%$ improvement in secrecy performance. Therefore, network engineers should prioritize the deployment of multi-RIS structures with larger RIS elements to secure the proposed RF-FSO model.

    \item  Fig. \ref{multi ris1} illustrates a significant reduction in SOP from $0.3151$ to $0.1649$ when increasing the number of RIS from $1$ to $3$, respectively, at the SNR of $10$ dB. This represents a $47.67\%$ improvement in secrecy performance, highlighting the effectiveness of deploying multiple RIS units to improve security.

    \item In Fig. \ref{turbulence}, at an SNR of $25$ dB, the SOP decreases from $0.0156$ to $0.0076$ when the pointing error parameter $\xi_{d}$ increases from $1.1$ to $6.7$ due to the strong turbulence conditions. This translates into a $51.28\%$ improvement in secrecy performance. To mitigate the effects of pointing errors, engineers should consider using larger optical receiver apertures, thereby improving the resilience of the system to minor pointing deviations.
    
    \item As depicted in Figure \ref{det}, the numerical analysis shows a significant reduction in SOP from $0.1373$ to $0.0156$ when $r_d$ is decreased from $2$ to $1$. This represents an improvement of $88.64\%$ in secrecy performance. Consequently, implementing HD techniques on the receiver is highly advantageous in enhancing the secrecy performance of the proposed model.
    
    \item In scenarios where there is a dominant LOS path, such as urban or open-field deployments, using a Rician fading model is recommended to ensure optimal performance.
\end{itemize}

\section{Conclusions} \label{con}

In this paper, we investigated the secrecy performance of a dual-hop multi-RIS-assisted RF-FSO communication system, analyzing the effects of various system parameters such as fading conditions, pointing errors, water salinity, RIS elements, and atmospheric turbulence. Through closed-form expressions and asymptotic analysis, key metrics including SOP, ASC, and EST were derived and validated using Monte Carlo simulations. Our numerical results show that secrecy performance improves with increased Rician fading parameters, owing to the stronger LOS component, while a higher eavesdropper fading severity degrades SOP performance. Additionally, increasing the number of RIS-reflecting elements significantly enhances signal focusing, leading to improved overall secrecy performance. Furthermore, the implementation of multiple RIS units provides additional diversity, ensuring stronger signal reception at the legitimate receiver and thus improving security performance. These findings underscore the potential of multi-RIS structures in enhancing the robustness of RF-FSO systems in environments susceptible to eavesdropping.

\bibliographystyle{IEEEtran}
\bibliography{IEEEabrv,asmbBiblio.bib}

\begin{thebibliography}{10}
\providecommand{\url}[1]{#1}
\csname url@samestyle\endcsname
\providecommand{\newblock}{\relax}
\providecommand{\bibinfo}[2]{#2}
\providecommand{\BIBentrySTDinterwordspacing}{\spaceskip=0pt\relax}
\providecommand{\BIBentryALTinterwordstretchfactor}{4}
\providecommand{\BIBentryALTinterwordspacing}{\spaceskip=\fontdimen2\font plus
\BIBentryALTinterwordstretchfactor\fontdimen3\font minus \fontdimen4\font\relax}
\providecommand{\BIBforeignlanguage}[2]{{%
\expandafter\ifx\csname l@#1\endcsname\relax
\typeout{** WARNING: IEEEtran.bst: No hyphenation pattern has been}%
\typeout{** loaded for the language `#1'. Using the pattern for}%
\typeout{** the default language instead.}%
\else
\language=\csname l@#1\endcsname
\fi
#2}}
\providecommand{\BIBdecl}{\relax}
\BIBdecl

\bibitem{shi2023outage}
Z.~Shi, H.~Wang, Y.~Fu, X.~Ye, G.~Yang, and S.~Ma, ``Outage performance and aoi minimization of harq-ir-ris aided iot networks,'' \emph{IEEE Transactions on Communications}, vol.~71, no.~3, pp. 1740--1754, 2023.

\bibitem{basharat2022reconfigurable}
S.~Basharat, S.~A. Hassan, A.~Mahmood, Z.~Ding, and M.~Gidlund, ``Reconfigurable intelligent surface-assisted backscatter communication: A new frontier for enabling 6g iot networks,'' \emph{IEEE Wireless Communications}, vol.~29, no.~6, pp. 96--103, 2022.

\bibitem{RAKIB2024}
M.~A. Rakib, M.~Ibrahim, A.~Badrudduza, I.~S. Ansari, M.~S.~U. Zaman, and H.~Yu, ``Ris-aided free-space optics communications in {A2G} networks over inverted gamma–gamma turbulent channels,'' \emph{ICT Express}, 2024.

\bibitem{han2020cooperative}
Y.~Han, S.~Zhang, L.~Duan, and R.~Zhang, ``Cooperative double-irs aided communication: Beamforming design and power scaling,'' \emph{IEEE Wireless Communications Letters}, vol.~9, no.~8, pp. 1206--1210, 2020.

\bibitem{tasci2022new}
R.~A. Tasci, F.~Kilinc, E.~Basar, and G.~C. Alexandropoulos, ``A new {RIS} architecture with a single power amplifier: Energy efficiency and error performance analysis,'' \emph{IEEE Access}, vol.~10, pp. 44\,804--44\,815, 2022.

\bibitem{peng2022performance}
Z.~Peng, X.~Chen, C.~Pan, M.~Elkashlan, and J.~Wang, ``Performance analysis and optimization for ris-assisted multi-user massive mimo systems with imperfect hardware,'' \emph{IEEE Transactions on Vehicular Technology}, vol.~71, no.~11, pp. 11\,786--11\,802, 2022.

\bibitem{rakib2024ris}
M.~A. Rakib, M.~Ibrahim, A.~Badrudduza, I.~S. Ansari, S.~Chakravarty, I.~Ahmed, and S.~A. Razzak, ``A ris empowered thz-uwo relay system for air-to-underwater mixed network: Performance analysis with pointing errors,'' \emph{IEEE Internet of Things Journal}, 2024.

\bibitem{sun2021performance}
Q.~Sun, Z.~Zhang, Y.~Zhang, M.~L{\'o}pez-Ben{\'\i}tez, and J.~Zhang, ``Performance analysis of dual-hop wireless systems over mixed fso/rf fading channel,'' \emph{IEEE Access}, vol.~9, pp. 85\,529--85\,542, 2021.

\bibitem{qu2022uav}
L.~Qu, G.~Xu, Z.~Zeng, N.~Zhang, and Q.~Zhang, ``Uav-assisted rf/fso relay system for space-air-ground integrated network: A performance analysis,'' \emph{IEEE Transactions on Wireless Communications}, vol.~21, no.~8, pp. 6211--6225, 2022.

\bibitem{ding2022performance}
J.~Ding, X.~Xie, L.~Wang, L.~Tan, J.~Ma, and D.~Kang, ``Performance of dual-hop fso/rf systems with fixed-gain relaying over fisher--snedecor f and $\kappa$-$\mu$ shadowed fading channels,'' \emph{Applied Optics}, vol.~61, no.~8, pp. 2079--2088, 2022.

\bibitem{ding2023joint}
J.~Ding, D.~Kang, X.~Xie, L.~Wang, L.~Tan, and J.~Ma, ``Joint effects of co-channel interferences and pointing errors on dual-hop mixed rf/fso fixed-gain and variable-gain relaying systems,'' \emph{IEEE Photonics Journal}, vol.~15, no.~1, pp. 1--11, 2023.

\bibitem{ashrafzadeh2019framework}
B.~Ashrafzadeh, E.~Soleimani-Nasab, M.~Kamandar, and M.~Uysal, ``A framework on the performance analysis of dual-hop mixed fso-rf cooperative systems,'' \emph{IEEE Transactions on Communications}, vol.~67, no.~7, pp. 4939--4954, 2019.

\bibitem{tonk2020mixed}
V.~K. Tonk, A.~Upadhya, P.~K. Yadav, and V.~K. Dwivedi, ``Mixed mud-rf/fso two way dcode and forward relaying networks in the presence of co-channel interference,'' \emph{Optics Communications}, vol. 464, p. 125415, 2020.

\bibitem{singya2020performance}
P.~K. Singya, N.~Kumar, V.~Bhatia, and M.-S. Alouini, ``On the performance analysis of higher order qam schemes over mixed rf/fso systems,'' \emph{IEEE Transactions on Vehicular Technology}, vol.~69, no.~7, pp. 7366--7378, 2020.

\bibitem{kong2021ergodic}
H.~Kong, M.~Lin, Z.~Wang, J.~Ouyang, and J.~Cheng, ``Ergodic capacity of high throughput satellite systems with mixed fso-rf transmission,'' \emph{IEEE Wireless Communications Letters}, vol.~10, no.~8, pp. 1732--1736, 2021.

\bibitem{sun2024performance}
Q.~Sun, Q.~Hu, Y.~Wu, X.~Chen, J.~Zhang, and M.~L{\'o}pez-Ben{\'\i}tez, ``Performance analysis of mixed fso/rf system for satellite-terrestrial relay network,'' \emph{IEEE Transactions on Vehicular Technology}, 2024.

\bibitem{liang2024performance}
J.~Liang, M.~Chen, and X.~Ke, ``Performance analysis of hybrid fso/rf-thz relay communication system,'' \emph{IEEE Photonics Journal}, 2024.

\bibitem{khalid2023reconfigurable}
W.~Khalid, M.~A.~U. Rehman, T.~Van~Chien, Z.~Kaleem, H.~Lee, and H.~Yu, ``Reconfigurable intelligent surface for physical layer security in 6g-iot: Designs, issues, and advances,'' \emph{IEEE Internet of Things Journal}, vol.~11, no.~2, pp. 3599--3613, 2023.

\bibitem{bhowal2022ris}
A.~Bhowal and S.~A{\"\i}ssa, ``Ris-aided communications in indoor and outdoor environments: Performance analysis with a realistic channel model,'' \emph{IEEE Transactions on Vehicular Technology}, vol.~71, no.~12, pp. 13\,356--13\,360, 2022.

\bibitem{zhang2024performance}
B.~Zhang, K.~Yang, K.~Wang, and G.~Zhang, ``Performance analysis for ris-assisted swipt-enabled iot systems,'' \emph{IEEE Transactions on Wireless Communications}, 2024.

\bibitem{atapattu2020reconfigurable}
S.~Atapattu, R.~Fan, P.~Dharmawansa, G.~Wang, J.~Evans, and T.~A. Tsiftsis, ``Reconfigurable intelligent surface assisted two--way communications: Performance analysis and optimization,'' \emph{IEEE Transactions on Communications}, vol.~68, no.~10, pp. 6552--6567, 2020.

\bibitem{guo2020outage}
C.~Guo, Y.~Cui, F.~Yang, and L.~Ding, ``Outage probability analysis and minimization in intelligent reflecting surface-assisted miso systems,'' \emph{IEEE Communications Letters}, vol.~24, no.~7, pp. 1563--1567, 2020.

\bibitem{selimis2021performance}
D.~Selimis, K.~P. Peppas, G.~C. Alexandropoulos, and F.~I. Lazarakis, ``On the performance analysis of ris-empowered communications over nakagami-m fading,'' \emph{IEEE Communications Letters}, vol.~25, no.~7, pp. 2191--2195, 2021.

\bibitem{basu2024performance}
A.~Basu, S.~P. Dash, A.~Kaushik, D.~Ghose, M.~Di~Renzo, and Y.~C. Eldar, ``Performance analysis of ris-aided index modulation with greedy detection over rician fading channels,'' \emph{IEEE Transactions on Wireless Communications}, 2024.

\bibitem{zhu2023ris}
X.~Zhu, L.~Yuan, Q.~Li, L.~Jin, X.~Nie, C.~Pan, and J.~Zhang, ``Ris-assisted full-duplex space shift keying: System scheme and performance analysis,'' \emph{IEEE Transactions on Green Communications and Networking}, 2023.

\bibitem{9057633}
L.~Yang, W.~Guo, and I.~S. Ansari, ``Mixed dual-hop fso-rf communication systems through reconfigurable intelligent surface,'' \emph{IEEE Communications Letters}, vol.~24, no.~7, pp. 1558--1562, 2020.

\bibitem{abualhayja2023exploiting}
M.~Abualhayja’a, A.~Centeno, L.~Mohjazi, M.~M. Butt, P.~Sehier, and M.~A. Imran, ``Exploiting multi-hop ris-assisted uav communications: Performance analysis,'' \emph{IEEE Communications Letters}, 2023.

\bibitem{9424709}
A.~M. Salhab and L.~Yang, ``Mixed rf/fso relay networks: Ris-equipped rf source vs ris-aided rf source,'' \emph{IEEE Wireless Communications Letters}, vol.~10, no.~8, pp. 1712--1716, 2021.

\bibitem{aldababsa2023multiple}
M.~Aldababsa, A.~M. Salhab, A.~A. Nasir, M.~H. Samuh, and D.~B. da~Costa, ``Multiple riss-aided networks: Performance analysis and optimization,'' \emph{IEEE Transactions on Vehicular Technology}, 2023.

\bibitem{10697101}
A.~B. Sarawar, A.~S.~M. Badrudduza, M.~Ibrahim, I.~S. Ansari, and H.~Yu, ``Secrecy performance analysis of integrated rf-uowc iot networks enabled by uav and underwater-ris,'' \emph{IEEE Internet of Things Journal}, pp. 1--1, 2024.

\bibitem{badrudduza2021security}
A.~Badrudduza, M.~Ibrahim, S.~R. Islam, M.~S. Hossen, M.~K. Kundu, I.~S. Ansari, and H.~Yu, ``Security at the physical layer over gg fading and megg turbulence induced rf-uowc mixed system,'' \emph{IEEE Access}, vol.~9, pp. 18\,123--18\,136, 2021.

\bibitem{shakir2021physical}
W.~M.~R. Shakir, ``Physical layer security performance analysis of hybrid fso/rf communication system,'' \emph{IEEE Access}, vol.~9, pp. 18\,948--18\,961, 2021.

\bibitem{mitev2023physical}
M.~Mitev, A.~Chorti, H.~V. Poor, and G.~P. Fettweis, ``What physical layer security can do for 6g security,'' \emph{IEEE Open Journal of Vehicular Technology}, vol.~4, pp. 375--388, 2023.

\bibitem{10460296}
M.~K. Ghosh, M.~Kumar~Kundu, M.~Ibrahim, A.~S.~M. Badrudduza, M.~S. Anower, I.~S. Ansari, A.~Solomon, S.~Chakravarty, I.~Ahmed, and H.~Yu, ``Physical-layer security in mixed uowc-rf networks with energy harvesting relay against multiple eavesdroppers,'' \emph{IEEE Open Journal of the Communications Society}, vol.~5, pp. 2884--2902, 2024.

\bibitem{saber2024security}
M.~J. Saber and M.~Hasna, ``Security analysis of integrated hap-based fso and uav-enabled rf downlink communications,'' \emph{IEEE Open Journal of the Communications Society}, 2024.

\bibitem{10058969}
M.~Ibrahim, A.~S.~M. Badrudduza, M.~S. Hossen, M.~K. Kundu, I.~S. Ansari, and I.~Ahmed, ``On effective secrecy throughput of underlay spectrum sharing $\alpha -\mu$/ málaga hybrid model under interference-and-transmit power constraints,'' \emph{IEEE Photonics Journal}, vol.~15, no.~2, pp. 1--13, 2023.

\bibitem{zhang2024joint}
Y.~Zhang, X.~Gao, H.~Yuan, K.~Yang, J.~Kang, P.~Wang, and D.~Niyato, ``Joint uav trajectory and power allocation with hybrid fso/rf for secure space-air-ground communications,'' \emph{IEEE Internet of Things Journal}, 2024.

\bibitem{yang2020secrecy}
L.~Yang, J.~Yang, W.~Xie, M.~O. Hasna, T.~Tsiftsis, and M.~Di~Renzo, ``Secrecy performance analysis of ris-aided wireless communication systems,'' \emph{IEEE Transactions on Vehicular Technology}, vol.~69, no.~10, pp. 12\,296--12\,300, 2020.

\bibitem{shi2024secrecy}
W.~Shi, J.~Xu, W.~Xu, C.~Yuen, A.~L. Swindlehurst, and C.~Zhao, ``On secrecy performance of ris-assisted miso systems over rician channels with spatially random eavesdroppers,'' \emph{IEEE Transactions on Wireless Communications}, 2024.

\bibitem{kaveh2023secrecy}
M.~Kaveh, Z.~Yan, and R.~J{\"a}ntti, ``Secrecy performance analysis of ris-aided smart grid communications,'' \emph{IEEE Transactions on Industrial Informatics}, 2023.

\bibitem{li2022enhancing}
X.~Li, Y.~Zheng, M.~Zeng, Y.~Liu, and O.~A. Dobre, ``Enhancing secrecy performance for star-ris noma networks,'' \emph{IEEE Transactions on Vehicular Technology}, vol.~72, no.~2, pp. 2684--2688, 2022.

\bibitem{zhang2022secrecy}
Z.~Zhang, J.~Chen, Y.~Liu, Q.~Wu, B.~He, and L.~Yang, ``On the secrecy design of star-ris assisted uplink noma networks,'' \emph{IEEE Transactions on Wireless Communications}, vol.~21, no.~12, pp. 11\,207--11\,221, 2022.

\bibitem{yadav2023secrecy}
A.~K. Yadav, S.~Yadav, A.~Pandey, and A.~Silva, ``On the secrecy performance of ris-enabled wireless communications over nakagami-m fading channels,'' \emph{ICT Express}, vol.~9, no.~3, pp. 452--458, 2023.

\bibitem{hoang2023secrecy}
T.~M. Hoang, C.~Xu, A.~Vahid, H.~D. Tuan, T.~Q. Duong, and L.~Hanzo, ``Secrecy-rate optimization of double ris-aided space--ground networks,'' \emph{IEEE Internet of Things Journal}, vol.~10, no.~15, pp. 13\,221--13\,234, 2023.

\bibitem{ruku2024effects}
M.~R.~A. Ruku, M.~Ibrahim, A.~Badrudduza, I.~S. Ansari, W.~Khalid, and H.~Yu, ``Effects of co-channel interference on ris empowered wireless networks amid multiple eavesdropping attempts,'' \emph{ICT Express}, vol.~10, no.~3, pp. 491--497, 2024.

\bibitem{wang2023uplink}
D.~Wang, M.~Wu, Z.~Wei, K.~Yu, L.~Min, and S.~Mumtaz, ``Uplink secrecy performance of ris-based rf/fso three-dimension heterogeneous networks,'' \emph{IEEE Transactions on Wireless Communications}, 2023.

\bibitem{10313311}
M.~M. Rahman, A.~S.~M. Badrudduza, N.~A. Sarker, M.~Ibrahim, I.~S. Ansari, and H.~Yu, ``Ris-aided mixed rf-fso wireless networks: Secrecy performance analysis with simultaneous eavesdropping,'' \emph{IEEE Access}, vol.~11, pp. 126\,507--126\,523, 2023.

\bibitem{ahmed2023enhancing}
T.~Ahmed, A.~Badrudduza, S.~R. Islam, S.~H. Islam, M.~Ibrahim, M.~Abdullah-Al-Wadud, and I.~S. Ansari, ``Enhancing physical layer secrecy performance for ris-assisted rf-fso mixed wireless system,'' \emph{IEEE Access}, vol.~11, pp. 127\,737--127\,753, 2023.

\bibitem{zhuang2022secrecy}
Y.~Zhuang and J.~Zhang, ``Secrecy performance analysis for a noma based fso-rf system with imperfect csi,'' \emph{Journal of Optical Communications and Networking}, vol.~14, no.~7, pp. 500--510, 2022.

\bibitem{gradshteyn1988tables}
I.~Gradshteyn, I.~Ryzhik, and R.~H. Romer, ``Tables of integrals, series, and products,'' 1988.

\bibitem{ibrahim2021enhancing}
M.~Ibrahim, A.~Badrudduza, M.~S. Hossen, M.~K. Kundu, and I.~S. Ansari, ``Enhancing security of tas/mrc-based mixed rf-uowc system with induced underwater turbulence effect,'' \emph{IEEE Systems Journal}, vol.~16, no.~4, pp. 5584--5595, 2021.

\bibitem{meijerfunction}
G.~Meijer, ``function: Integration(subsection 21/02/03/01), functions. wolfram. com.''

\bibitem{ansari2015performance}
I.~S. Ansari, F.~Yilmaz, and M.-S. Alouini, ``Performance analysis of free-space optical links over m{\'a}laga turbulence channels with pointing errors,'' \emph{IEEE Transactions on Wireless Communications}, vol.~15, no.~1, pp. 91--102, 2015.

\bibitem{art3}
A.~P. Prudnikov, A.~Brychkov, and O.~I. Marichev, \emph{Integrals and series: special functions}.\hskip 1em plus 0.5em minus 0.4em\relax CRC Press, 1986, vol.~2.

\end{thebibliography}

\end{document}